\documentclass{aa}  

\usepackage{graphicx}
\usepackage{txfonts}
\usepackage{natbib}
\usepackage{multirow}
\usepackage{rotating}
\usepackage{txfonts}
\usepackage{tabularray}
\usepackage{subcaption}
\usepackage{footnotehyper}
\usepackage{array}
\usepackage{booktabs}
\usepackage{caption}
\usepackage{subcaption}
\usepackage{gensymb}
\usepackage{amssymb}
\usepackage{booktabs}
\usepackage{url}
\usepackage{multirow}
\usepackage{siunitx}
\usepackage{wrapfig}
\usepackage{lscape}
\usepackage{rotating}
\usepackage{longtable}
\usepackage{epstopdf}
\usepackage{caption}
\usepackage{txfonts}
\usepackage{natbib}
\usepackage{subcaption}
\usepackage{longtable}
\usepackage{array}
\usepackage{caption}
\usepackage{tablefootnote}
\usepackage{soul}
\newcommand{\solar}{$_{\odot}$}
\newcommand{\micron}{$\mu$m}
\newcommand{\standform}[1]{$\times$10$^{#1}$}
\newcolumntype{y}[1]{>{\centering\let\newline\\\arraybackslash\hspace{0pt}}p{#1}}
\bibpunct{(}{)}{;}{a}{}{,} 
\begin{document} 

   \title{OGLE-2002-BLG-360: a dusty anomaly among red nova remnants}
   \author{T. Steinmetz\inst{1}
          \and
          T. Kami\'{n}ski\inst{1}
          \and
          C. Melis\inst{2}
          \and
          N. Blagorodnova \inst{3,4,5}
          \and
          M. Gromadzki\inst{6}
          \and
          K. Menten\inst{7}
          \and
          K. Su\inst{8}
          }

   \institute{Nicolaus Copernicus Astronomical Center, ul. Rabia\'{n}ska 8, 87-100 Toru\'{n}, Poland \email{[thomas,tomkam]@ncac.torun.pl}
   \and
   Astronomy \& Astrophysics Department, University of California, San Diego, La Jolla, CA 92093-0424, USA 
   \and 
   Institut de Ciències del Cosmos (ICCUB), Universitat de Barcelona (IEEC-UB), Martí Franquès 1, 08028, Barcelona, Spain 
   \and 
   Departament de Física Quàntica i Astrofísica (FQA), Universitat de Barcelona (UB), c. Martí i Franquès, 1, 08028, Barcelona, Spain 
   \and 
   Institut d'Estudis Espacials de Catalunya (IEEC), 08860, Castelldefels (Barcelona), Spain 
   \and
   Astronomical Observatory, University of Warsaw, Al. Ujazdowskie 4, PL-00-478 Warszawa, Poland 
   \and
   Max-Planck-Institut für Radioastronomie, Auf dem Hügel 69, D-53121 Bonn, Germany 
   \and
   Space Science Institute, 4750 Walnut Street, Suite 205, Boulder, CO 80301, USA 
   }

   \date{Received XXX, accepted YYY}

  \abstract
   {OGLE-2002-BLG-360 is an example of a Galactic red nova, the aftermath of a non-compact stellar merger. The dusty nature of the material surrounding the merger remnant makes observations of this particular source difficult, meaning the properties of the central star and its surrounding environment are poorly understood.}
   {We aim to establish the characteristics of the merger remnant, as well as of the dusty environment and its structure. We attempt to establish similarities with other Galactic red novae and argue how such an environment may have formed.}
   {We use infrared, and sub-millimeter observations to construct the spectral energy distribution (SED) between 2\,\micron\ and 1.27\,mm for an epoch 15--21\,years after the red nova eruption. We use the radiative transfer codes DUSTY and RADMC-3D to model the SED and retrieve the physical properties of both the central star and the surrounding dust.}
   {We show that the SED is best replicated by a spherically symmetric model consisting of an M-type supergiant surrounded by 0.012\,M\solar\ of dust concentrated within two spherical shells. The dust in the outer shell extends out to a maximum distance of 9500\,AU from the central source, whilst the inner shell extends out to 1000\,AU. The dust composition is dominated by iron grains (58\%), but also contains olivine silicates (25\%) and alumina dust (17\%), which are both required to reproduce the profile of the observed 10\,\micron\ absorption feature.}
   {The inner shell likely originates from merger and post-merger ejecta, whilst the outer shell consists of material lost much earlier, before the merger event occurred. Evolution of the SED indicates continued dust formation within the expanding inner shell, which may be analogous to winds of red supergiants. The object is extremely dusty compared to other Galactic red nova remnants.}

\maketitle


\section{Introduction}\label{sect: intro}
Red novae (also known as red transients) are a class of eruptive objects characterised by multi-peaked light curves \citep{metzger2017} and rapid cooling to low effective stellar temperatures \citep{tylenda2005,pastorello2019}.  The aftermath of red nova eruptions evolve into a cool giant star and are associated with significant dust formation, which gives rise to a luminous infrared (IR) remnant. Red novae have been suggested to be the result of stellar mergers between non-compact stars by \citet{soker2003} and \citet{tylenda2006}, based on the evolution of V838 Mon \citep[Nova 2002;][]{munari2002}, which is believed to have formed from the merger event of two main-sequence (MS) stars. Over the last two decades, however, it has become evident that red nova outbursts can also be powered by mergers of stars at other evolutionary stages, such as protostars, red giants, asymptotic giant branch stars, yellow supergiants, or white dwarfs \citep[see e.g.][]{stepien2011, blagorodnova2021, tylenda2024}. 

Since the eruption of V838 Mon in 2002, five more red novae have been recognised in the Milky Way, two of which were discovered prior to V838 Mon: CK Vul \citep[Nova 1670;][]{kato2003}, and V4332 Sgr \citep[Nova 1994;][]{martini1999}. The other three are V1309 Sco \citep[Nova 2008;][]{mason2010}, OGLE-2002-BLG-360 \citep[Nova 2002;][hereafter T13]{tylenda2013}, and ZTF SLRN-2020 \citep{De2023}. Multiple other red novae have been discovered in the Local Group of galaxies \citep[see][and references therein]{pastorello2019}. Several different sources, for which only the centuries-old remnants have been identified, have also been suggested as the result of mergers, such as the Blue Ring nebula \citep{hoadley2020}, as well as Phoenix Giant stars \citep{melis2009}. Several similarities exist between these sources and red novae, such as large amounts of circumstellar dust and evidence for past mass-loss events. However, these sources were not observed during the coalescence.

Red novae show many features commonly seen in evolved stars. The remnant stellar source cools down to temperatures of $\approx$3000 K, similar to late M-type stars \citep{kaminski2010v4332,nicholls2013}. For example, V838 Mon exhibited SiO maser emission \citep{ortiz2020}, as well as a thick dust-driven wind typical of red supergiants. Bipolar outflows have also been observed in several Galactic red novae. CK Vul has a clear hourglass-shaped nebulosity \citep{hajduk2007}, whilst bipolar outflows and circumstellar structures have been identified in V838 Mon \citep{mobeen2024}, V4332 Sgr \citep{kaminski2018}, and V1309 Sco \citep{steinmetz2024}.

OGLE-2002-BLG-360 (hereafter BLG-360) was discovered by the OGLE Early Warning System \citep[EWS][]{udalski2003} on 9 October 2002. The early light curve was analysed by \citet{paczynski2003} and wrongly identified as a long microlensing event involving a Galactic Bulge object. However, \citet{tylenda2013} recognized that the OGLE light curve displayed characteristics typical of other Galactic red novae. Firstly, the light curve clearly shows three peaks. The first of the three peaks, often referred to as `precursor' in the literature, is also considerably fainter than the other peaks, similar to V838 Mon. The remnant star cools from an M3 to M6-type star between 2003 and 2010, and shows a rise to peak luminosity characteristic of that seen during the common envelope phase in V1309 Sco. Therefore, BLG-360 bears many hallmarks of a red nova. Unfortunately, no spectral observations were obtained for the object during or shortly after the outburst.

BLG-360 is unique among red novae because it exhibited the longest outburst out of the known Galactic red novae. For example, whereas the three peaks seen in the light curve of V838 Mon spanned just three months, the three peaks seen in the light curve of BLG-360 spanned a total of almost 4 years.
The distance to BLG-360 was estimated by T13 to be 8.2 kpc, based on the OGLE-III extinction maps produced by \citet{nataf2013}. This estimate was regarded by the authors as a rough constraint and not a precise value. Based on the position of BLG-360, the value of $E(B-V)$ ranges between 0.89--1.10 mag, depending on the adopted value of R$_V$. Spectral energy distribution (SED) modelling by T13 gave the first (and only) measurements of the source properties. Based on data taken between March--September 2010 ($\approx$ 8 years after the beginning of the outburst), the central stellar remnant was a red supergiant of spectral type M6--7, with an effective temperature, T$_{\rm eff}$ of 3200 K and a radius of 300 R\solar. The luminosity of the star was estimated as 8200 L\solar, with an optically thick ($\tau_V \geq$ 20), warm (550 K) dust shell surrounding the source. This is considerably different to the estimated properties of the progenitor, where T13 determined that the progenitor was an M3 red giant primary with T$_{\rm eff}$=4300 K, a radius of 30 R\solar\ and luminosity of $\approx$300 L\solar. T13 found that the pre-outburst SED required obscuration by warm (800 K) dust, which provides an extinction (A$_V$) of 3 mag and indicated that steady dust formation occurred during the progenitor phase. If the progenitor was a low-mass giant, this level of extinction would agree with the lower estimates of the models from \citet{macleod2022} and so provide further evidence of a giant progenitor with pre-existing dust obscuration before the merger.

In this paper, we combine observations with a plethora of archival photometric data to construct the  SED of BLG-360 spanning between infrared and radio wavelengths and taken between 15--21 years after the precursor outburst. We use the radiative transfer codes DUSTY \citep{dustyv2} and RADMC-3D \citep{radmc3d} to examine both the stellar and dust properties in the circumstellar environment. In particular, we aim to examine the dust structure and composition to investigate the chemical processes that took place during the merger event. The dust spatial distribution may also give us some insight into the evolutionary history of the source before, during, and after the merger event.

The paper is organised into the following sections. Section \ref{sect: obs} describes the photometric and spectroscopic observations, Sect. \ref{sect: results} describes the results of the SED modelling. Section \ref{sect: discussion} discusses the implications on stellar and dust properties. Section \ref{sect: T13 results discussion} traces the evolution of the source, and Sect. \ref{sect: compare transients} discusses the unique properties of BLG-360. In Sect. \ref{sect: summary} we present our conclusions.

\begin{figure}[h!]
    \centering
    \includegraphics[width=\linewidth]{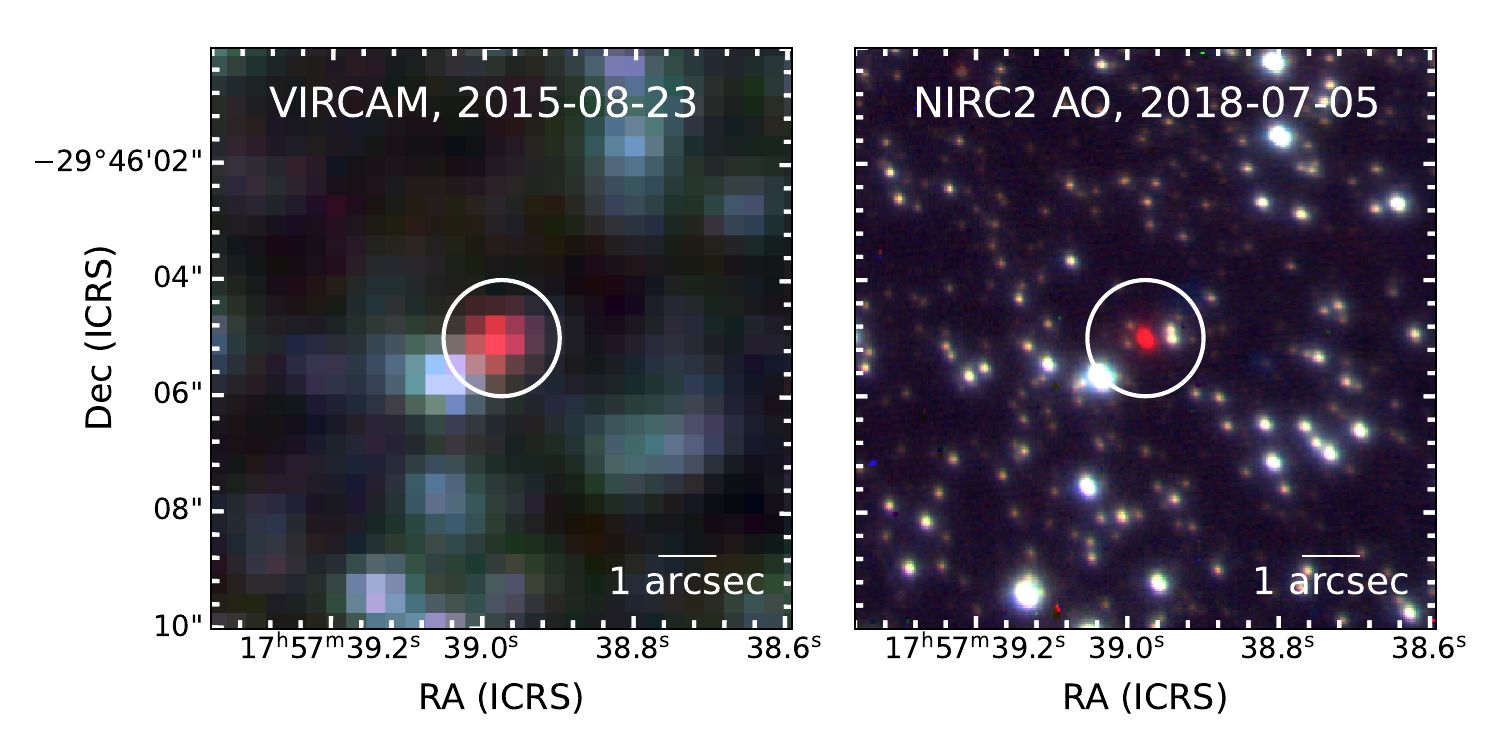}
    \caption{Colour composite of $J$ (blue), $H$ (green), $Ks$ (red) bands of the field of OGLE-360 observed in two different epochs with VIRCAM/VISTA (left) and NIRC2\,AO/Keck (right). The white circle indicates the position of OGLE-360 with a 1\arcsec\ radius. The AO image shows nearby sources that can potentially interfere with the object's photometry.}
    \label{fig-nirc2}
\end{figure}

\section{Observations}\label{sect: obs}
In this section we describe the various observations of BLG-360, in order of increasing wavelength, used to construct the SED, including non-detections. The photometric fluxes are listed in Table \ref{tab: SED photometry}. When collecting archival data, we selected observations or measurements at epochs close to our SOFIA observations. In order to get the best possible wavelength coverage, we select data within four years of the SOFIA observations.

\begin{figure*}[h!]
    \centering
    \includegraphics[scale=0.35]{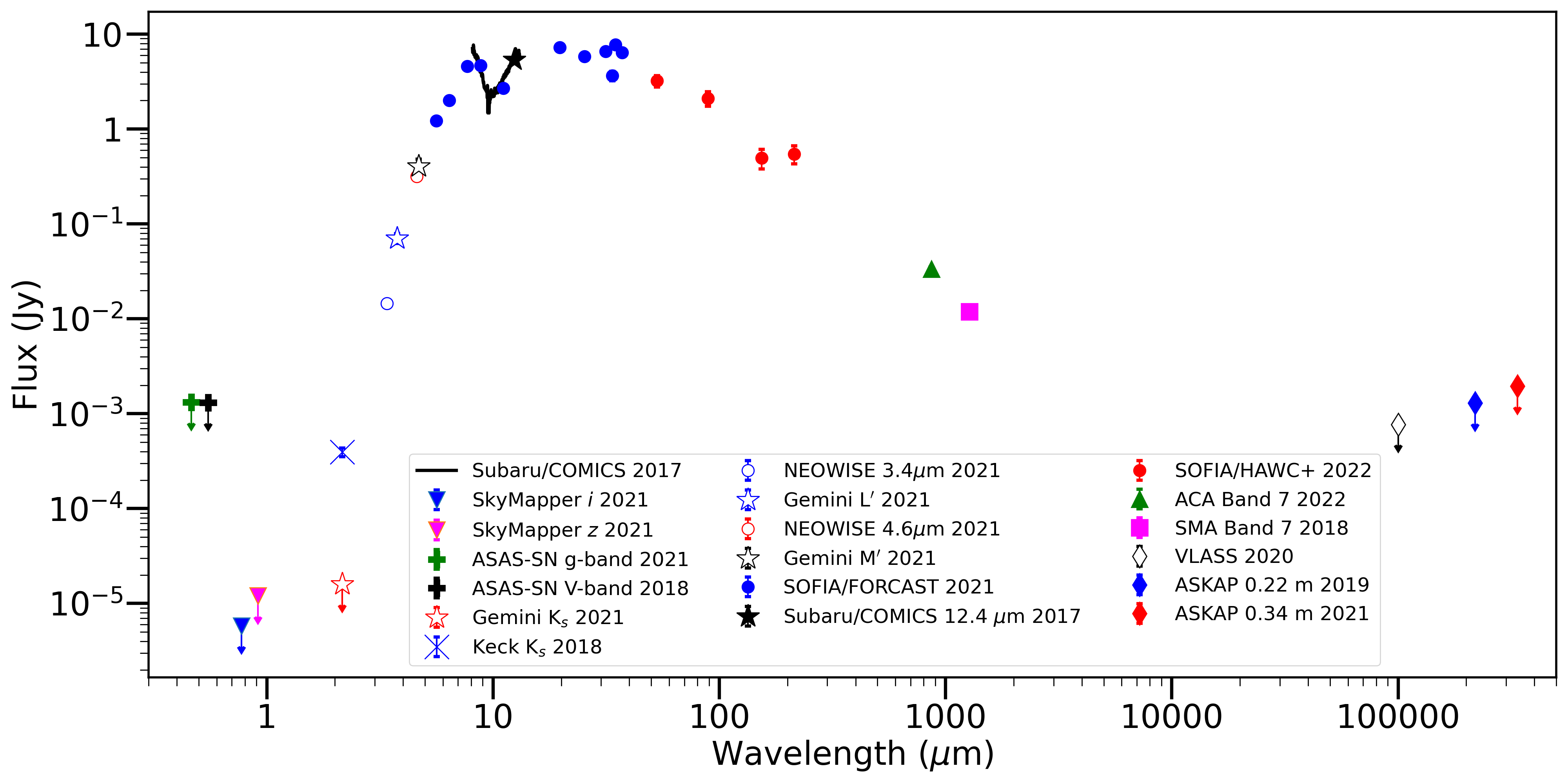}
    \caption{Observed SED of BLG-360 between 2017--2022. 5$\sigma$ upper limits from ASAS-SN, ASKAP, and VLASS are indicated by arrows pointed downwards, and flux measurements from detections show standard errors. All data points have associated errors, and for those where the errorbars cannot be seen, the errors are within the area covered by the markers. }
    \label{fig: SED}
\end{figure*}

\subsection{Keck}\label{sect: keck}
BLG-360 was observed using the Natural Guide Star Adaptative Optics (NGS-AO) available for the near-infrared (NIR) imager \citep[NIRC2;][]{nirc2_2000PASP} on the Keck II telescope located at Maunakea in Hawaii. The observations took place on 05/07/2018 UT with the $J$ (1.248\,\micron), $H$ (1.633\,\micron), and $K_s$ (2.146\,\micron) bands. The approximate airmass was 1.6. The observations were obtained in wide camera mode, covering a field of view (FoV) of 40\arcsec$\times$40\arcsec\ with 0\farcs04/pixel. The exposure times were 15\,s with 4 co-adds for all filters.

The data were reduced using custom-developed \texttt{python} routines. These included calibration using master darks and flats obtained as part of the daytime calibrations. The contribution from the sky background was removed from each exposure using a median-combined "empty" sky region obtained just after the target observations at a similar airmass and sky area. The astrometry of each field was calibrated using the position of stars from the Gaia DR3 catalogue \citep{GaiaDR32023AA} that were detectable in each image.

The photometry of the sources was obtained using aperture photometry with an aperture radius of 3.0$\times$ the seeing in each filter. The zeropoints were derived by comparing the photometry of non-blended stars in the field of view with the measurements available in the VVV DR2 catalogue \citep{VVVDR2_Minniti_2017yCat} using the corrected aperture magnitude with 2\arcsec\ diameter. BLG-360 was detected in the $K_s$ band across three frames, but not in the $J$ or $H$ bands. The extracted $K_s$ magnitudes of 15.6$\pm$0.1, 15.6$\pm$0.1 and 15.5$\pm$0.1 mag were converted to fluxes using the $K_s$ filter information provided by the Spanish virtual observatory (SVO\footnote{http://svo2.cab.inta-csic.es/theory/fps/}). The final flux, taken as an average over the three frames, is 0.39 $\pm$ 0.04 mJy. The three-colour $JHK_s$ image is shown in Fig. \ref{fig-nirc2}. The 5$\sigma$ non-detection limits derived for the $H$ and $J$ bands are 21.1 and 20.5\,mag respectively.

\subsection{Gemini}\label{sect: gemini}
BLG-360 was observed using the near-infrared (NIR) imager \citep[NIRI;][]{niri} on the Gemini North telescope at Maunakea in Hawaii. The observations took place on 19/10/2021 with the $K_s$ (2.15 \micron) and $L^\prime$ (3.77 \micron) bands, and on 21/10/2021 with the $M^\prime$ (4.68 \micron) band. Five science images were obtained for all three bands, followed by five sky images, using the f/32 camera. Due to detector issues during the initial part of the observations, the first 1--2 frames taken were omitted for each band. The exposure times used were 10, 0.8, and 0.2 s for $K_s$, $L^\prime$, and $M^\prime$, with 4, 30, and 70 co-added frames, respectively, resulting in total exposure times per image of 40, 24, and 16 s for the three respective bands. The observed airmass in all three bands ranged between 2.00--2.16, and the observational conditions in the $K_s$ and $L^\prime$ bands were clear or photometric.

The source is well-detected in the $L^\prime$ and $M^\prime$ bands with magnitudes of 8.85 $\pm$ 0.05 and 6.50 $\pm$ 0.08 for $L^\prime$ and $M^\prime$, respectively. The situation is far less clear for the $K_s$ band image; after careful comparison to the $L^\prime$ image and the Keck/NIRC2 AO images we conclude that we cannot reliably identify BLG-360 in the Gemini $K_s$ band data. While some weak emission appears to be present around the expected source position (see Fig. \ref{fig-nirc2}), it is not clear if this emission is from BLG-360, some background sources, or a combination thereof. Given the confusion-limited regime the $K_s$ band data appear to be in, we extract the flux around the expected position for the target source with an aperture comparable to that used for the flux calibration source and adopt this value as an upper limit. The extracted counts correspond to a $K_s$ band magnitude of 19.0. Using the Vega magnitude system and zero point fluxes from SVO, we obtained fluxes of $<$ 1.6\standform{-5} Jy, 70.7 $\pm$ 8.1, and 410 $\pm$ 75 mJy for the $K_s$, $L^\prime$, and $M^\prime$ bands, respectively.

\subsection{(NEO)WISE}\label{sect: wise}
\begin{figure}
    \centering
    \includegraphics[scale=0.35]{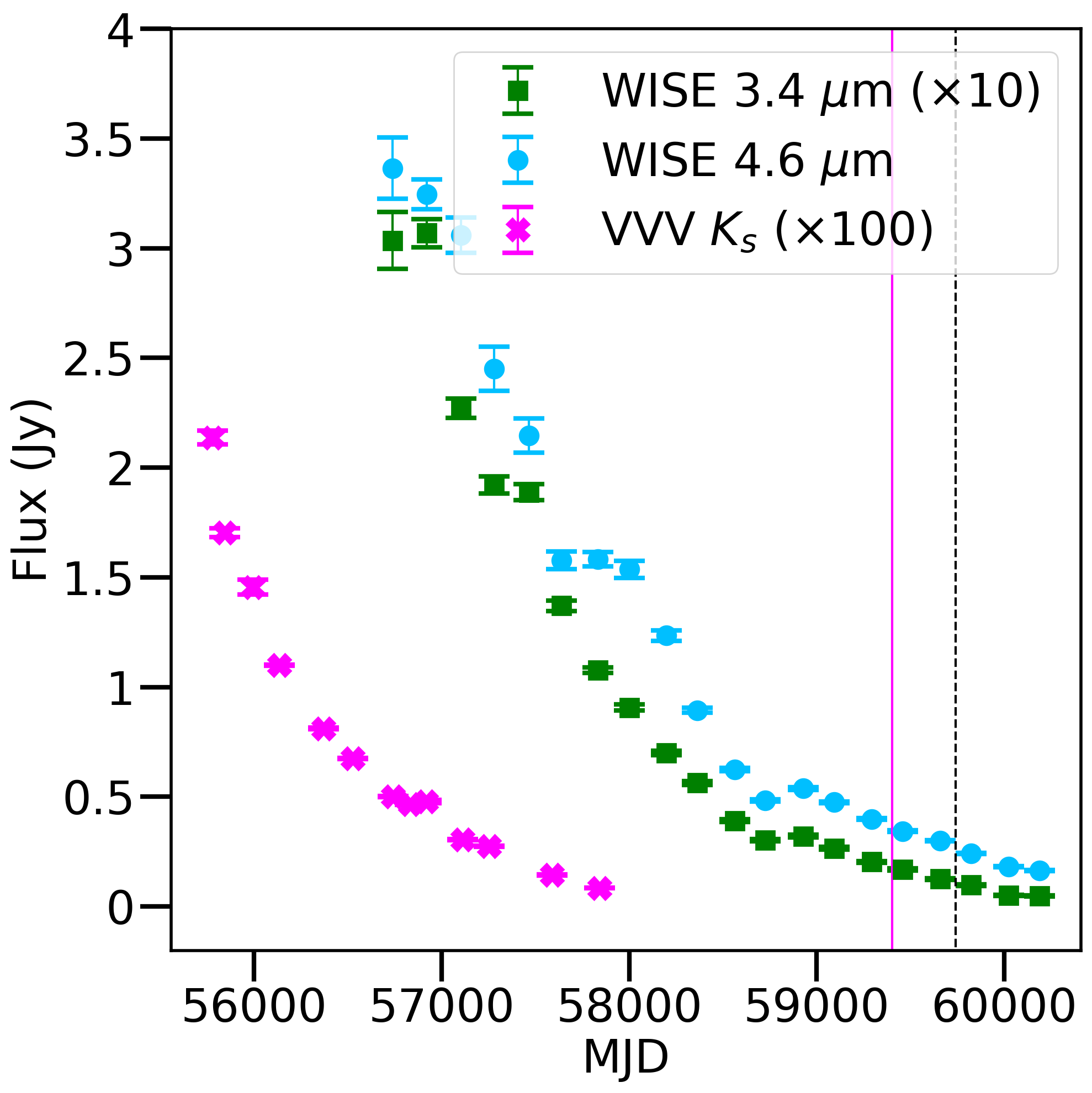}
    \caption{Light curve of BLG-360 at 3.4 \micron\ (green) and 4.6 \micron\ (blue), observed with NEOWISE between 2014--2023. The 3.4 \micron\ light curve is scaled up by a factor of 10. Also included is the VVV $K_s$ light curve at 2.15 \micron\, extracted between 2010--2016, and scaled up by a factor of 100. The magenta and black vertical lines indicate the epochs for the SOFIA FORCAST and HAWC+ observations, respectively.}
    \label{fig: neowise lc}
\end{figure}

BLG-360 was observed by the NEOWISE \citep{neowise} survey at multiple epochs between 2014--2023. The observations were obtained through the NEOWISE-R Single Exposure (L1b) Source Table in the NEOWISE reactivation database, and included BLG-360 data at 19 separate epochs, in the 3.4 and 4.6 \micron\ bands. The NEOWISE fluxes were measured regularly, approximately every 6 months from March 2014 to September 2023. To construct the light curve, we filtered out any measurements which did not match the following criteria: the data quality must be graded AA, the signal-to-noise (SNR) > 2, and the instrumental profile-fit photometry reduced $\chi^2$ must be less than 150, to ensure that adjustments to the PSF are reasonable. The NEOWISE light curve (Fig. \ref{fig: neowise lc}) shows an overall decline in both filters between 2014--2023. The 3.4 \micron\ flux decreased linearly until MJD$\approx$58700, after which the rate of decline of the flux decreases significantly and the light curve plateaus. Plateaus are also seen at MJD$\sim$57300 and MJD$\sim$58000 for the 3.4 \micron\ and 4.6 \micron\ light curves, respectively. A slight rise in both fluxes is also seen at MJD$\sim$58900. The cause of the rise and plateaus are unclear, but may be related to episodic dust formation as well as possible selection effects regarding the filters used to neglect the `bad' NEOWISE data from the binned light curve.

In order to determine a single flux for the NEOWISE filters to include within our SED, we first identified the NEOWISE fluxes from epochs at the same time as the SOFIA observations with both FORCAST and HAWC+. During these times, the slope in both NEOWISE filters appeared approximately linear, and so we interpolated the flux at the midpoint between the NEOWISE observations in both filters to determine the flux at the midpoint between the FORCAST and HAWC+ observations (MJD 59571). The interpolated fluxes are 0.019 $\pm$ 0.008 and 0.27 $\pm$ 0.091 Jy for 3.4 and 4.6 \micron, respectively. These fluxes are presented in Table \ref{tab: SED photometry}. We did not apply a colour correction to the measurements, as the correction is rather uncertain. Since the blue part of the observed SED is best approximated by a Planck function at a temperature of 400 K, the corrections would be 1.13 and 1.02 at 3.4 and 4.6 \micron, respectively \citep{WISEcolorcorr}. These are insignificant considering other uncertainties.
\begin{table}
    \centering
    \small
    \caption{Photometry, including non-detections, used to construct the SED of BLG-360.}
    \begin{tabular}{ccccc}\hline
MJD & Instrument & Filter & $\lambda_{\rm c}$ & Flux$^1$\\
& & & (\micron) & (Jy)\\\hline
59564 & ASAS-SN & $g$ & 0.46 & < 0.0013\\
58270 & ASAS-SN & $V$ & 0.55 & < 0.0013\\
59421 & SkyMapper & $i$ & 0.77 & < 5.8\standform{-6}\\
59421 & SkyMapper & $z$ & 0.91 & < 1.2\standform{-5}\\
58304 &  Keck-NIRC2 & $K_s$ & 2.15 & (3.94 $\pm$ 0.40)\standform{-4} \\
59506 & Gemini+NIRI & $K_s$ & 2.15 & < 1.6\standform{-5} \\
59571 & NEOWISE$^2$ & W1 & 3.40 & (1.46 $\pm$ 0.03)\standform{-2} \\
59506 & Gemini+NIRI & $L^\prime$ & 3.77 & 0.071 $\pm$ 0.008 \\
59571 & NEOWISE$^2$ & W2 & 4.60 & 0.32 $\pm$ 0.020 \\
59508 & Gemini+NIRI & $M^\prime$ & 4.68 & 0.41 $\pm$ 0.075 \\
59403 & FORCAST & F056 & 5.61 & 1.22 $\pm$ 0.09 \\
59403 & FORCAST & F064 & 6.35 & 2.00 $\pm$ 0.13 \\
59403 & FORCAST & F077 & 7.72 & 4.60 $\pm$ 0.27 \\
59403 & FORCAST & F088 & 8.64  & 4.68 $\pm$ 0.28 \\
59403 & FORCAST & F111 & 11.01 & 2.69 $\pm$ 0.18 \\
57935 & COMICS & N12.4 & 12.40 & 5.43 $\pm$ 0.28 \\
59403 & FORCAST & F197 & 19.70 & 7.29 $\pm$ 0.44 \\
59403 & FORCAST & F253 & 25.24 & 5.81 $\pm$ 0.37 \\
59403 & FORCAST & F315 & 31.36 & 6.58 $\pm$ 0.44 \\
59403 & FORCAST & F336 & 33.57 & 3.65 $\pm$ 0.41 \\
59403 & FORCAST & F348 & 34.64 & 7.76 $\pm$ 0.52 \\
59403 & FORCAST & F371 & 36.98 & 6.44 $\pm$ 0.43 \\
59739 & HAWC+ & Band A & 53.0 & 3.22 $\pm$ 0.45 \\
59739 & HAWC+ & Band C & 89.0 & 2.11 $\pm$ 0.46 \\
59739 & HAWC+ & Band D & 154.0 & 0.49 $\pm$ 0.13 \\
59739 & HAWC+ & Band E & 214.0 & 0.67 $\pm$ 0.15 \\
59715 & ACA & Band 7 & 867 & 0.034 $\pm$ 0.001 \\
58304 & SMA & 230 GHz & 1273 & 0.012 $\pm$ 0.005 \\
59154 & VLASS & 3 GHz & 10$^5$ & < 7.72$\times$10$^{-4}$\\
58795 & ASKAP & 1.37 GHz & 2.19e5 & < 1.30$\times$10$^{-3}$\\
59228 & ASKAP & 0.89 GHz & 3.37e5 & < 1.95$\times$10$^{-3}$\\\hline
    \end{tabular}
    \tablefoot{$^1$ Non-detection fluxes are given as 5$\sigma$ upper limits, where $\sigma$ is the root-mean-square noise in a region where BLG-360 should be observed. We refer to the flux errors as 1$\sigma$ uncertainties.\\
    $^2$ These fluxes were obtained using interpolation between observations epochs to match the MJD mid-point between the SOFIA FORCAST and HAWC+ observations epochs.}
    \label{tab: SED photometry}
\end{table}
\subsection{Subaru}\label{sect: subaru}
BLG-360 was observed on 01/07/2017 using the COMICS  \citep{comics1,comics2} imager and spectrograph on the Subaru telescope situated at Maunakea. The Subaru data are the earliest dataset in our SED. Both spectroscopic and imaging observations were taken simultaneously, and implemented a chopping throw of 10$^{\prime\prime}$ with a chopping frequency of 0.5 Hz in order to reduce the high background levels. Imaging was taken using the N12.4 filter ($\lambda_c$=12.4 \micron, $\Delta\lambda$=1.2 \micron) with a total exposure time of 0.005--0.04 s per exposure, with ten frames being co-added into a single frame. The plate scale of the detector is 0.13\arcsec/pixel with FoV of 42\arcsec$\times$32\arcsec. The photometric flux measured was 5.43 $\pm$ 0.28 Jy, although this error is likely underestimated.

\subsection{SOFIA}\label{sect: sofia}
We utilised the Stratospheric Observatory for Infrared Astronomy (SOFIA) facility for our mid to far-infrared (mid-IR) observations. We used both the FORCAST \citep{forcast} and HAWC+ \citep{hawk+} instruments to obtain photometry between 5--214 \micron, and describe the observations below.
\subsubsection{FORCAST}\label{sect: forcast}
The FORCAST imaging observations were obtained on 08/07/2021 during SOFIA flight 756 in nod-match-chop mode with a chop throw of 75\arcsec\ and a chop angle of 110\degree. Three-point dithering was applied. FORCAST observations were done in the dual configuration of the instrument -- that is, in two filters simultaneously -- and a total of eleven filters were used. The exposure times range from 30 to 120 s, depending on the filter combination. The raw data were processed and calibrated in the FORCAST Redux pipeline version 2.3.0 \citep{sofiaredux}.

Aperture photometry was performed on level 3 pipeline products using large apertures of 12 pixels. Fluxes and errors were measured using the FORCAST photometry script provided by the SOFIA Observatory\footnote{https://github.com/SOFIAObservatory/Recipes/blob/master/\\FORCAST-Basic\_Photometry.ipynb}. Uncertainties took into account calibration errors and pixel flux variations. The latter dominate the uncertainties as the FORCAST images are not flat field corrected. No colour correction was applied, as it is of the order of 1\% or less, and so does not improve significantly the accuracy of our measurements. Table 1 lists central wavelengths of filters used. BLG-360 is the brightest point source in the field but two other sources were marginally detected 40\arcsec\ west of BLG-360 and at wavelengths shorter than 9\,\micron.

To examine the fluctuations in the FORCAST photometry, we extracted fluxes using PSF fitting, following the methodology of \citet{burris2023}. The results can be found in Appendix \ref{appendix: burris photometry}. The results show that the fluctuations can be explained by noise within the standard 12-pixel extraction aperture, but as the new FORCAST photometry do not impact our modelling results, we do not include the alternative photometry in our models fits.

\subsubsection{HAWC+}\label{sect: hawc}
BLG-360 was observed on 09/06/2022 with HAWC+ during flight 884 in on-the-fly mapping (OTFMAP) scan mode, using the Lissajous scan pattern without any chopping or nodding. The observations were taken in bands A, C, D, and E (53, 89, 154, and 214 \micron). Band B was not used as it has a known over-saturation issue described in the HAWC+ documentation\footnote{https://irsa.ipac.caltech.edu/data/SOFIA/docs/instruments/\\handbooks/HAWC\_Handbook\_for\_Archive\_Users\_Ver1.0.pdf}. Exposure times for bands A, C, D, and E were 93, 950, 253, and 140 s respectively. The half-wave plate was kept open throughout observations.

The data were reduced using the HAWC DRP pipeline v3.0.0 \citep{sofiaredux}. The fluxes and errors were extracted using a modified version of the photometry script used for FORCAST. Errors may be slightly underestimated as the calibration errors and pixel flux variations are less certain for HAWC+ compared to FORCAST. The source was detected in all four bands, with fluxes and errors shown in Table \ref{tab: SED photometry}. 

Spectroscopy was carried out with the G01L10L grism using an exposure time per frame of 0.3 s, with 3 exposures co-added into a single frame. The spectrum was acquired using the NL (N band, low resolution) grating with a dispersion of 0.02 \micron/pixel. It covers 8.1--13.1 \micron, showing the 10 \micron\ absorption band.

\subsection{ALMA/ACA}\label{sect: aca}
BLG-360 was observed on 16/05/2022 using the Atacama Compact Array (ACA, a.k.a. Morita Array) consisting of 7-m antennas. Ten antennas were used in total. Band 7 data were taken between 342.14--346.08 GHz and 354.12--358.05 GHz, with a velocity resolution of $\approx$1 km s$^{-1}$. The total integration time was $\approx$1400 s, the maximum recoverable scale was 19\farcs5, and the FoV was 28\farcs5. The data were reduced using CASA pipeline 6.2.1.7. The integrated continuum flux is 33.7 $\pm$ 1.11 mJy. A source size of (835$\pm$501) $\times$ (372$\pm$210) mas was calculated, deconvolved from a clean beam of 4\farcs84 $\times$ 2\farcs79 in natural weighting. This deconvolved size is likely just an upper limit. The heterodyne setup covers the rest frequency ($\nu_{\rm rest}$=345.796 GHz) of CO (3--2), but no lines were detected towards BLG-360. By measuring the spatial root-mean-square (rms; denoted as $\sigma$) of the background in 200 km s$^{-1}$ bins, we get a 5$\sigma$ upper limit on the circumstellar CO flux of 13.5 mJy at the fitted position of BLG-360 ($\alpha$=17$^h$ 57$^m$ 38.9796$^s$, $\delta$=--29$\degree$ 46\arcmin 04\farcs952). We do, however, detect very narrow interstellar emission 18\arcsec\ north-east from the position of BLG-360. This cloud is at the edge of the ACA's FoV (see Sect. \ref{sect: sma} below for more details about the interstellar emission region). 

\subsection{SMA}\label{sect: sma}
BLG-360 was observed in two runs, the first executed on 2/7/2018 and 7/7/2018, and again between 24/5/2019 and 26/5/2019 with the Submillimeter Array (SMA) at Maunakea. The 2018 observations used 7 antennas on each run. The 2019 observations were taken using 8 antennas on 24/5/2019, 7 antennas on 25/5/2019 and 6 antennas on 26/5/2019. 

The observations were carried out in the compact array configuration, with baselines of up to 70 m. The gain calibrators used for 2018 were 1700--261, 1733--130, 1924--292, 3C279, 3C454.3, 2253+161, and MWC349A whilst the flux calibrators used were Mars, Neptune, and Venus. For the 2019 observations, the phase calibrators were 1256--057, 1700--261, 1733--130, 1751+096, and 3C279. The only flux calibrator used was Callisto. The first run taken on 02/07/2018 was flagged as unsatisfactory by the observer and so is not included in this analysis.

\begin{figure}
    \centering
    \includegraphics[trim=20 0 10 40, scale=0.65, clip]{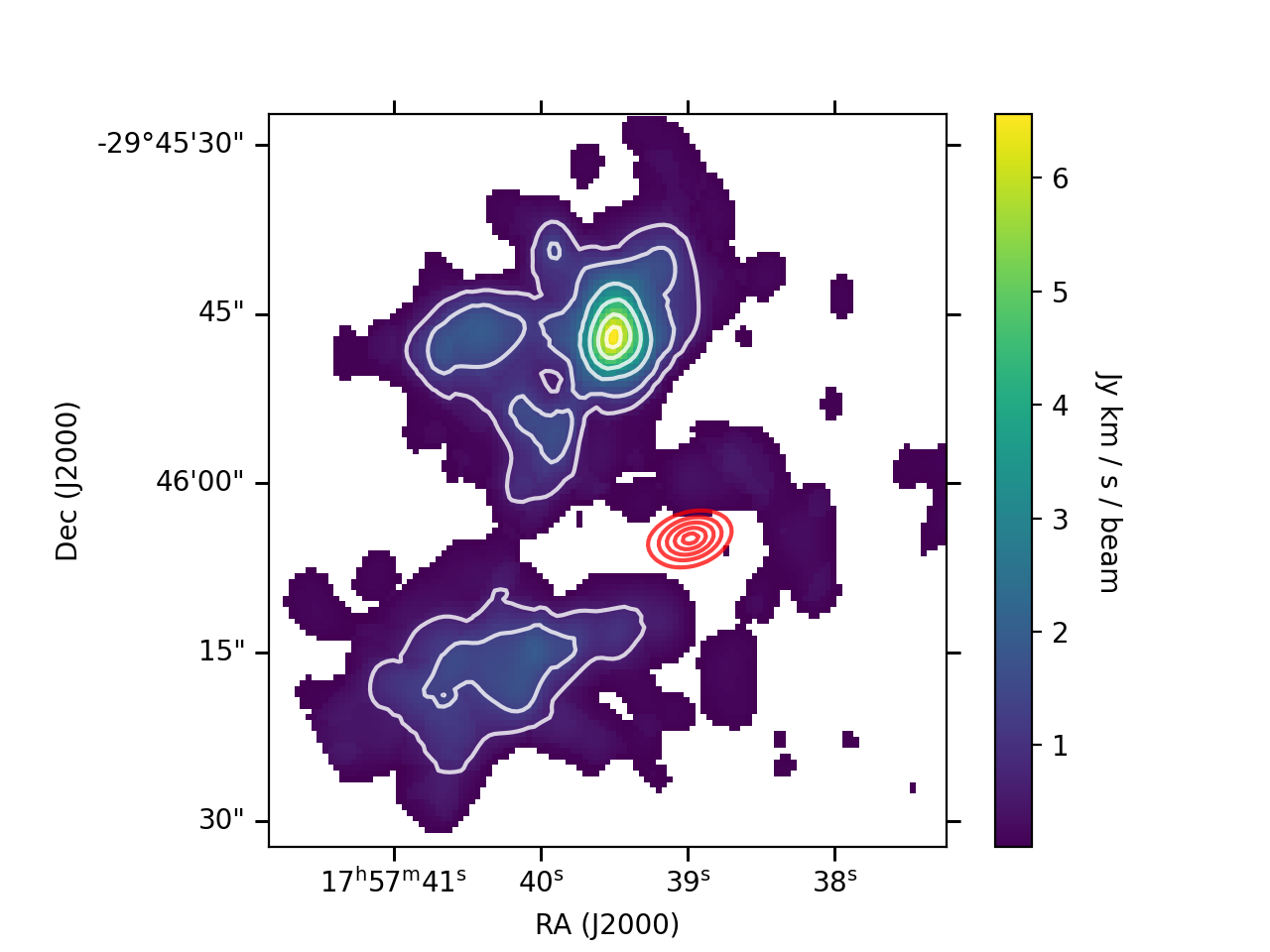}
    \caption{Integrated intensity of CO (2--1) from SMA overlaid with white contours at 10, 20, 40, 60, 80, and 95\% of the peak intensity. The ACA continuum is plotted in red contours at 20, 40, 60, 80, and 95\% of the ACA peak intensity. The ACA continuum is indicative of the source. }

    \label{fig: sma+aca}
\end{figure}

The SMA heterodyne receivers were configured so that the lower sideband covered 223.4--231.6 GHz (1294--1342 \micron), whilst the upper sideband covered 239.4--247.6 GHz (1211--1252 \micron). The source is only detected in continuum at the fitted position $\alpha$=17$^h$ 57$^m$ 38.9779$^s$, $\delta$=--29$\degree$ 46\arcmin 05\farcs4279, although interstellar CO (2--1) emission is seen at extended (up to 36\arcsec) angular distances, with a strong peak at $\varv_{\rm LSR}$=17 km s$^{-1}$ at $\alpha$=17$^h$ 57$^m$ 39\fs41, $\delta$=--29$\degree$45\arcmin 46\farcs71 (see Fig. \ref{fig: sma+aca}). This is likely an unassociated molecular cloud. As can be seen in Fig. \ref{fig: sma+aca}, BLG-360 is located in a CO cavity, that is, no CO emission overlaps with the position of the red nova. The emission of CO (3--2) observed with the ACA coincides with the most intense interstellar cloud observed in CO (2--1) and is centered at the same LSR velocity. 

The clean beam size was 4\farcs24 $\times$ 2\farcs82, with a beam PA of 170$\degree$ in natural weighting. The deconvolved source size is, as an upper limit, 1\farcs3 $\times$ 0\farcs85. The continuum flux density of BLG-360 is 11.9 $\pm$ 0.45 mJy/beam at 235.5 GHz.

\subsection{Non-detections and upper limits}\label{sect: upper limits}
\paragraph{\textbf{ASAS-SN}}
Using the ASAS-SN Sky Patrol photometry database\footnote{https://asas-sn.osu.edu/} \citep{shappee2014,kochanek2017}, we extracted photometric light curves at the location of BLG-360 in the $g$ (0.46 \micron) and $V$ (0.55 \micron) bands observed during epochs close to the SOFIA epochs. In the case of the $V$ band, we averaged the photometric data taken between 08/2/2018 and 22/9/2018. For the $g$ band, we considered two observations taken on 3/2/2021 and 27/10/2022.

The fluxes shown in the SED (Fig. \ref{fig: SED}) are upper limits, due to the fact that the pixel size of 8\arcsec\space and PSF of 15\arcsec\space for ASAS-SN means that it is likely that these fluxes are contaminated by nearby field stars. These upper limits are given as 1.31 mJy for the $V$ band, and 1.32 mJy for the $g$ band, corresponding to magnitudes of 16.12 and 16.20, respectively, in the Vega magnitude system.

\paragraph{\textbf{SkyMapper}}
We used the public cutout service for the DR4 of the SkyMapper survey \citep{SkyMapperDR42024PAS} to obtain images for the field of BLG-360 in the $i$ and $z$ bands. The images that were taken the closest in time to the SOFIA/FORCAST measurements correspond to MJD 59421.4. Although the median seeing was 3\farcs7 and 3\farcs1 respectively, the depth and smaller pixel size (0\farcs5) represent an improvement over the ASASS-SN optical dataset. Using the zeropoints provided by the survey, we derive a 5$\sigma$ non-detection limits of $i=$22.0 and $z=$21.2\,mag in the AB system, corresponding to 5.8 $\mu$Jy and 12.0 $\mu$Jy.

\paragraph{\textbf{VLA}}
BLG-360 was observed at three separate epochs on 15/02/2018, 01/11/2020, and 15/06/2023 by the Karl G. Jansky Very Large Array (VLA) as part of the VLA Sky Survey \citep[VLASS;][]{vlass}. The archival data were downloaded directly as reduced multi-frequency synthesis (mfs) images from the archive (single combined image across a range of frequencies). All data were observed as part of observations of the J175817-293000 field. The VLASS survey implemented the BnA configuration for all of these observations. Each observation consists of a single mfs image centred at 3 GHz ($\lambda$=10 cm). The angular resolutions range from 2\farcs1--2\farcs4. BLG-360 is covered but not detected in these observations, but 5$\sigma$ upper limits on the BLG-360 flux at this frequency were extracted from a region with a radius of 5\arcsec\space centred at the J2000 coordinates of BLG-360. These are 658, 772, and 440 $\mu$Jy in 2018, 2020 and 2023 respectively. As the observations in 2020 are closest to the SOFIA observation epoch, we adopt 772 $\mu$Jy as the VLA flux upper limit.

\paragraph{\textbf{ASKAP}}
Two fields containing BLG-360 by the Australian Square Kilometre Array Pathfinder (ASKAP) were found using the CASDA archive service\footnote{https://data.csiro.au/domain/casda}. These observations were taken on 8/11/2019 at 0.89 GHz and on 14/1/2021 at 1.37 GHz, and are cutouts from the RACS\_1752-231A and RACS\_1748-28 fields, respectively. The restoring beams for the 0.89 GHz and 1.37 Hz observations are 16\farcs2 $\times$ 14\farcs04, with a position angle of 102$\degree$, and 8\farcs64 $\times$ 7\farcs2, with a position angle of 69$\degree$, respectively. The 2019 observation was reduced using the ASKAP pipeline 1.0-RC-11267 and further processed using CASA v5.3.0-143.e17 and BPTOOL. The 2021 observation was reduced using the ASKAP pipeline v1.9.9, and subsequently processed using CASA v6.2, BPTOOL v2.4, and CONVOLVED v2.0.4.

BLG-360 was covered but was not detected at rms noise levels of 0.39 and 0.26 mJy for the 0.89 and 1.37 GHz observations, respectively.  We use 5$\sigma$ values as our upper limits. 
 The rms noise levels were measured using a region with a radius of 20\arcsec\space centred on the coordinates of BLG-360. No source is seen within this aperture. The SED, comprising of all aforementioned data (Sects. \ref{sect: keck}--\ref{sect: upper limits}), is shown in Fig. \ref{fig: SED}.

\section{Results}\label{sect: results}
\subsection{Extinction and reddening}\label{sect: reddening}
T13 adopted $E(B-V)$=1.0 mag based on the reddening of planetary nebulae within one square degree of the source. We revise $E(B-V)$ using reddening of Mira stars within 0.8$\degree$ of BLG-360 from \citet{lewis2023}. A total of 100 Miras sources were used, which show little variation in estimated extinction, with a standard deviation of 0.17 mag. The mean value is A$_{Ks}$ = 0.67 $\pm$ 0.17 mag. Using the standard values of A$_{Ks}$/A$_V$=0.118 and R$_V$=A$_V$/$E(B-V)$=3.1 \citep{ccm89}, we find $E(B-V)$=1.83$\pm$0.46 mag using the Miras technique. This is not significantly higher than the reddening adopted by T13. Applying a more recent extinction law given by \citet{nishiyama2008} for the Galactic Center, where A$_{Ks}$/A$_V$=0.062$\pm$0.005 mag and R$_V$=1.8, we find that $E(B-V)$=6.00 $\pm$1.60 mag, which is a factor of $\approx$3--4 larger than the range considered in T13. We subsequently advocate a smaller distance to BLG-360, locating it well outside the Galactic Center. Therefore, we adopt the $E(B-V)$ value of 1.83 mag, derived from nearby Miras, for our analysis. The total extinction-corrected flux of BLG-360 is then $F=\int F_{\lambda} d\lambda=3.14\times$10$^{-9}$ erg s$^{-1}$ cm$^{-2}$ (3.14$\times$10$^{-12}$ W m$^{-2}$). This is equivalent to a bolometric luminosity of $6.6 \times 10^3 \times d^2/(8.2\,{\rm kpc})^2$ L\solar\ at the T13 distance.
\subsection{Spectral energy distribution}\label{sect: blg360 properties}
The SED shows a steady rise in flux between 2--8 \micron\ that can be approximated by a linear gradient of log$_{10}$(flux)/log$_{10}$($\lambda$ (\micron))$\approx$7.3, followed by a plateau between 8--37 \micron\ with strong 10 \micron\ silicate absorption fully resolved in the Subaru spectrum. At $\lambda >$ 19.7 \micron, the slope changes to --0.85, although fluctuations in the FORCAST fluxes mean that this is not a perfect linear fit. The SED between 89--1273 \micron\ is best represented by a steeper spectrum with a spectral index of --1.86. We therefore see a `knee' in the Rayleigh-Jeans tail of the SED between 53--154 \micron, although the data coverage in this wavelength range is sparse.

\subsection{SED modelling}
Our initial analysis consisted of fitting the observed SED using the 1-D radiative transfer code DUSTY \citep{dustyv2}. This analysis gave a good visual fit to the SED  (see Fig. \ref{fig: model D}) by implementing a single, spherically symmetric shell of dust. The dust shell, however, extended out to 20 pc from the central star, which is inconsistent with typical circumstellar sizes of evolved stars \citep{fong2006}. The method, results, and analysis of our DUSTY modelling is given in Appendix \ref{appendix: dusty}.

To get a closer physical representation of the SED, we used RADMC-3D\footnote{https://github.com/dullemond/radmc3d-2.0} \citep{radmc3d} to manually fit the SED. To start off, we consider only spherically symmetric geometries in our modelling (this is discussed later in Sect.\,\ref{sect: sphere v disk}). We examined three different dust structures: a single shell of continuous dust (S), a two-shell structure (D), and the same D geometry but with a cavity with zero dust density between the inner and outer dust shell (CD). We treat the central star as a point source, which is the only heat source in the system.

\begin{figure}
    \centering
    \includegraphics[trim={0 97 100 100},clip=True,width=\columnwidth]{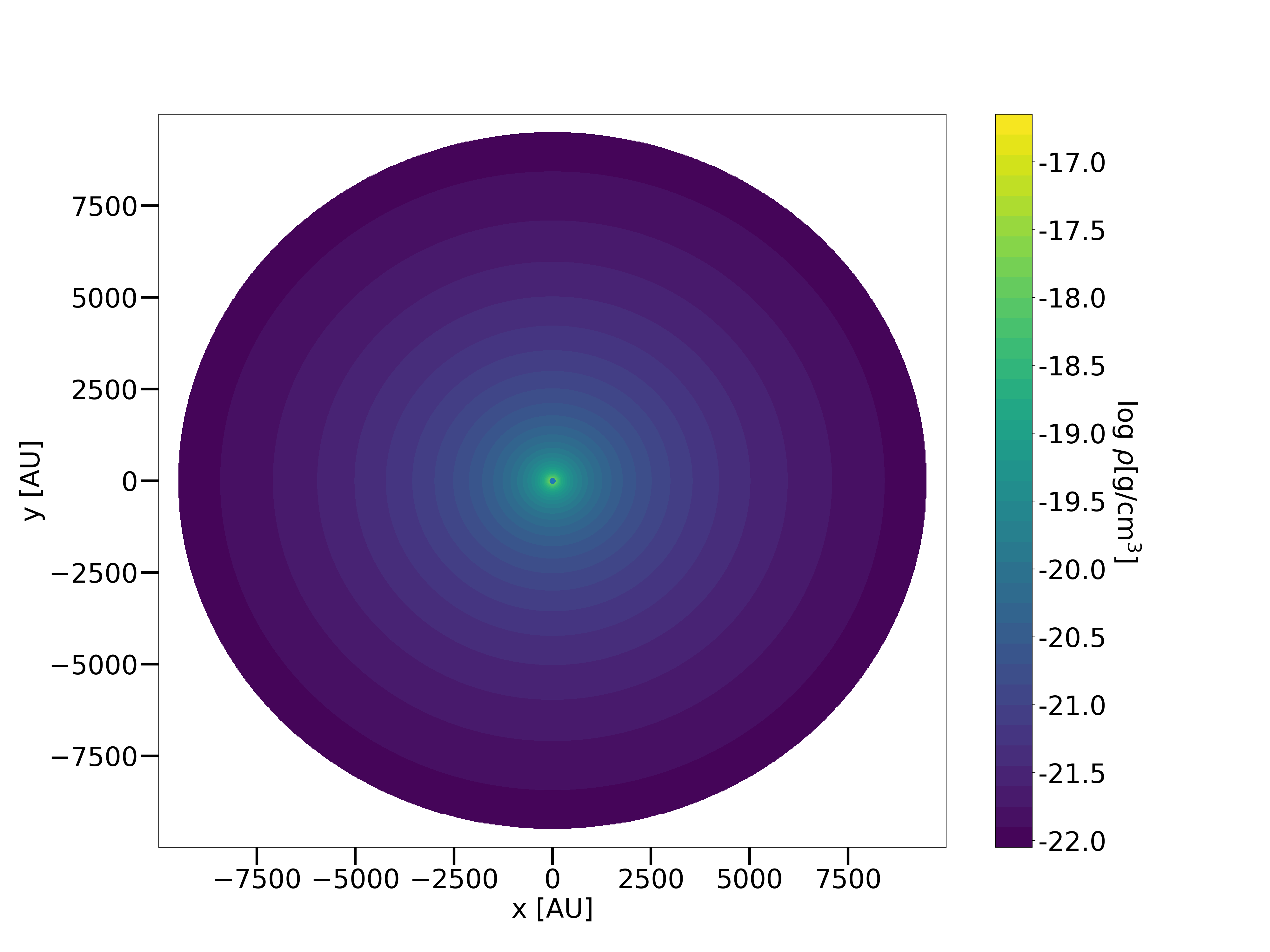}
    \includegraphics[trim={0 97 100 100},clip=True,width=\columnwidth]{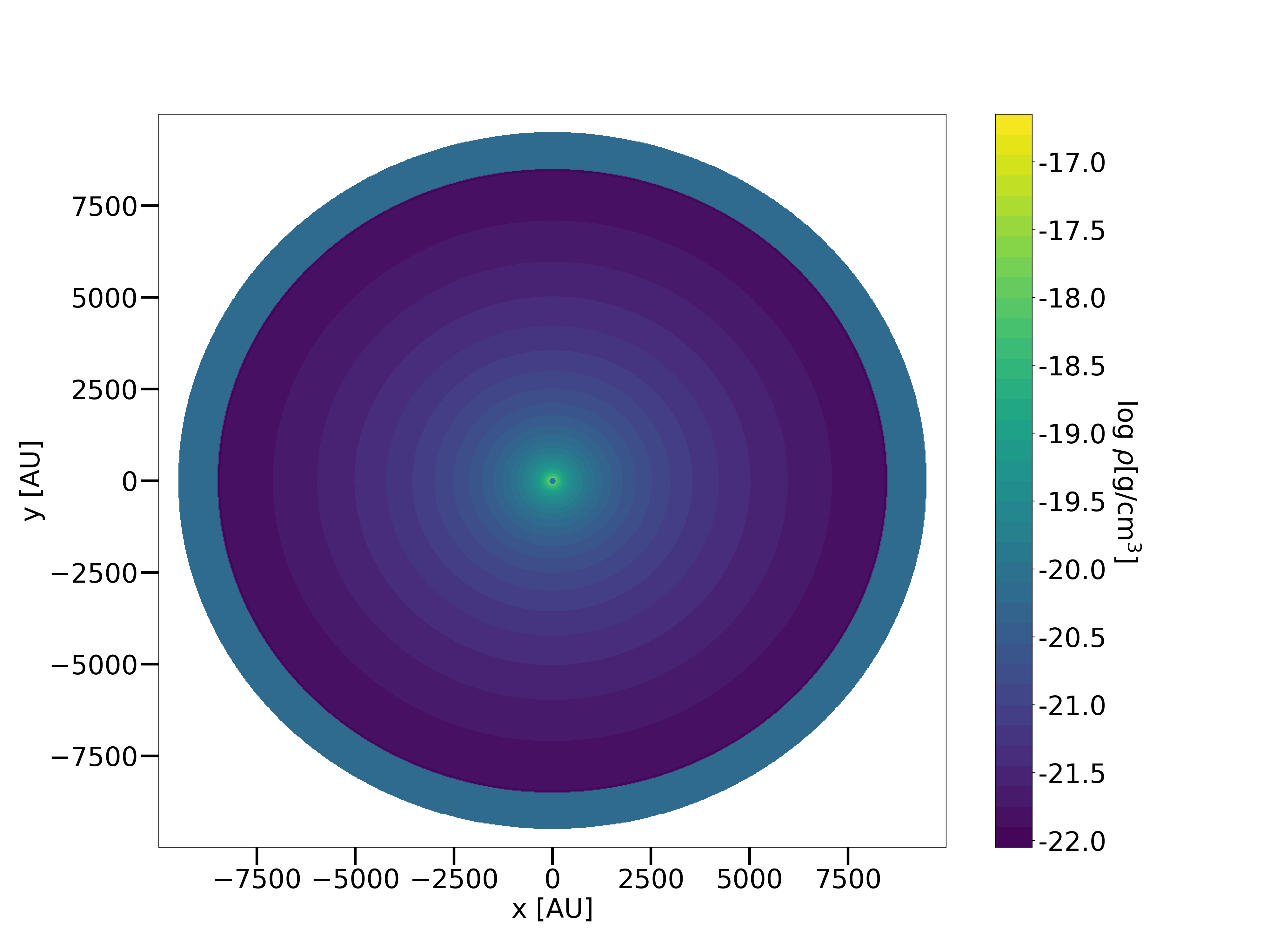}
    \includegraphics[trim={0 35 100 100},clip=True,width=\columnwidth]{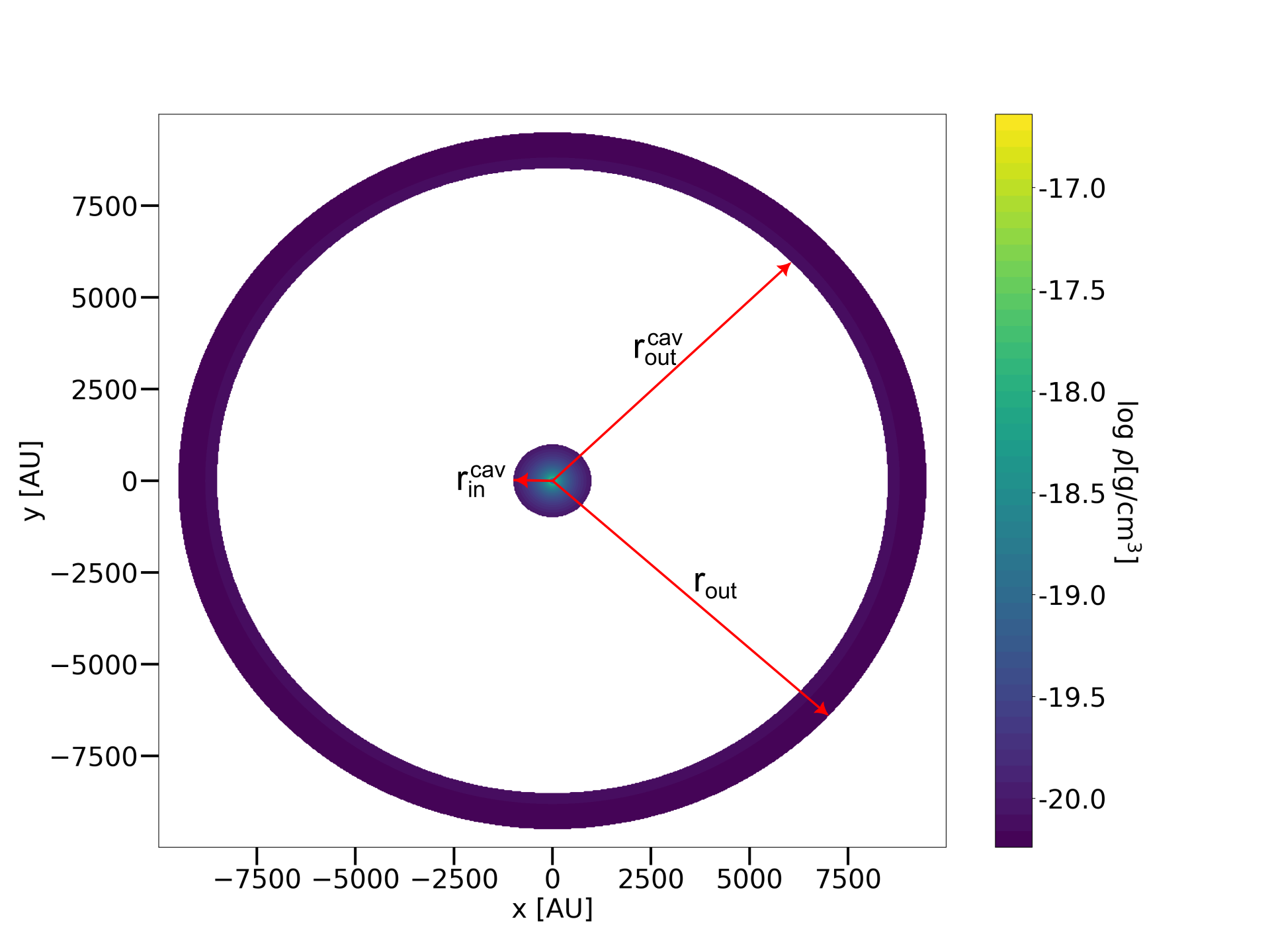}
    \caption{Spherically symmetric dust geometries used in RADMC-3D, centred on the star. Colourbars show the logarithm of dust density in g cm$^{-3}$. \textbf{Top}: Single continuous shell of dust (S). \textbf{Middle}: Continuous dust shell with an overdense shell at outer edge (D). \textbf{Bottom}: Two-shell model with cavity (CD). The labelled arrows indicate the size parameters used in the model, except for r$_{\rm in}$, which is unresolved at this resolution.}
    \label{fig: radmc3d model geometries}
\end{figure}

Dust distributions for these models are depicted in Fig. \ref{fig: radmc3d model geometries}. Our models contained silicates, alumina dust \citep{begemann1997}, and solid iron dust \citep{pollack1994,henningstognienko1996}. We considered several silicate types, and found olivine grains \citep{dorschner1995olivinepyroxene} to best replicate our observations. We also examined possible contributions from FeO \citep{henning1995}, MgS \citep{begemann1994}, and amorphous carbon \citep{preibisch1993} grains but find these do not fit our observations. We mix the grains using Mie theory via python routines supplied in the RADMC-3D package. For the silicate dust, we only implement one grain type, and do not mix different silicate grains together in any single model. We do this for simplicity, as investigating a mix of silicate grains introduces additional degrees of freedom.

Our models are calculated across box grids spanning from --10000 to 10000 AU in the x, y, and z-axes, centred on the star. We use a grid resolution of 800 points on each axis, giving a spatial resolution of 25 AU per grid cell. The star is represented by a black-body with a Planck spectrum and is parametrised by the temperature, T$_{eff}$, and the radius, R$_{\star}$. We implement the dust distribution through several parameters, including the maximal density $\rho_{max}$; inner and outer radii of the dust shells (r$_{\rm in}$ and r$_{\rm out}$) and cavity (r$_{\rm in}^{\rm cav}$ and r$_{\rm out}^{\rm cav}$); and density scaling factor (f$_p$) for the outer shell, which is applied to dust at radii greater than the boundary radius r$_{\rm bound}$. For our CD models, r$_{\rm out}^{\rm cav}$=r$_{\rm bound}$, as the outer shell in our models is formed from scaling the density of the outer dust at r $\geq\, r_{\rm bound}$. We fix the dust density gradient p$_{\rho}$ ($\rho \propto\ $r$^{p_{\rho}}$) to 2.00, as larger values increase the optical depth to an extent where the calculation time increases significantly, whilst lower values result in insufficient dust to obscure the star. We set the dust sublimation temperature arbitrarily at 1700 K, and remove any dust that exceeds this temperature from the model. 

\begin{figure*}
    \centering
    \includegraphics[scale=0.2]{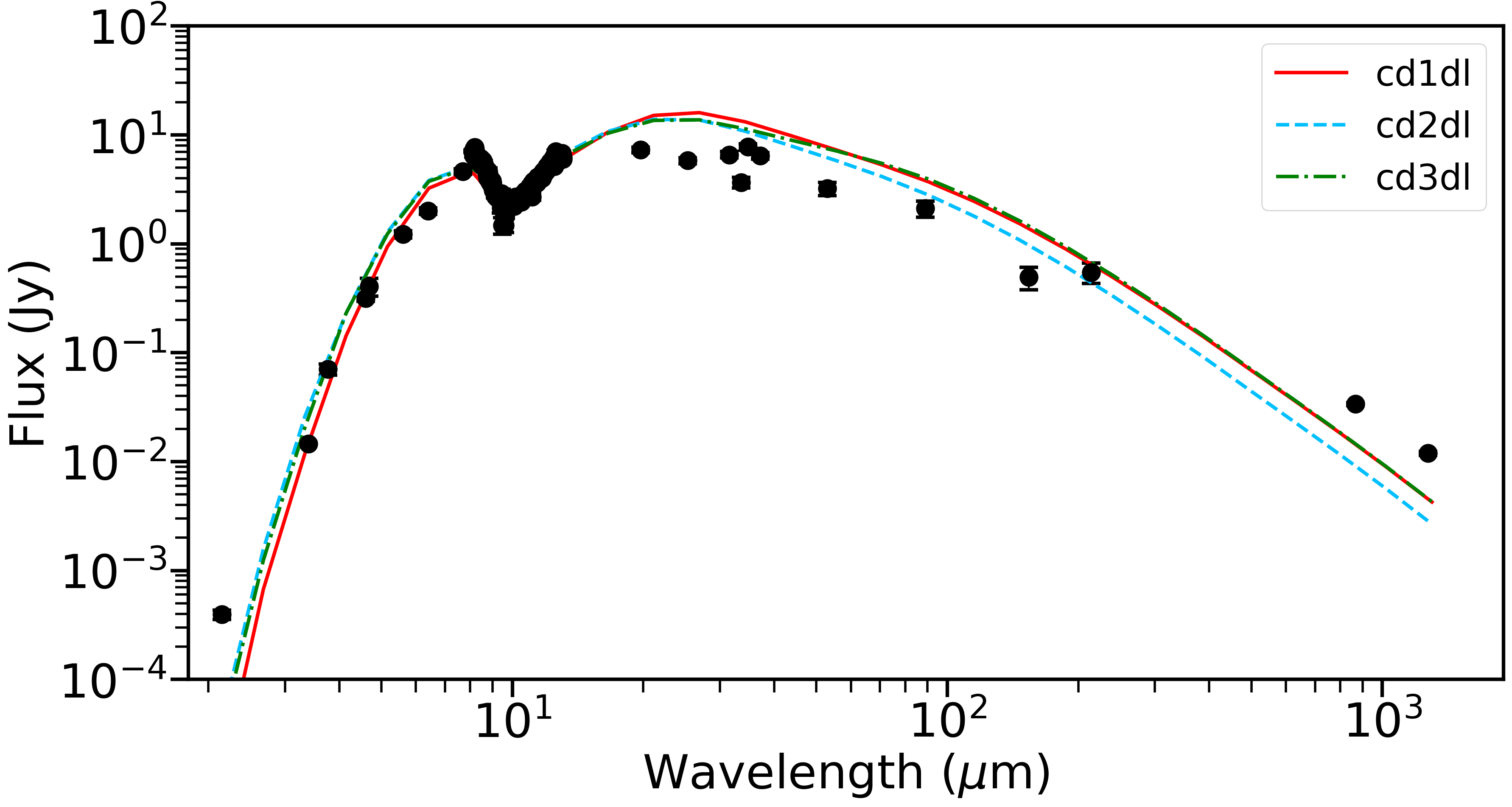}
    \vspace{0.2mm}
    \includegraphics[scale=0.2]{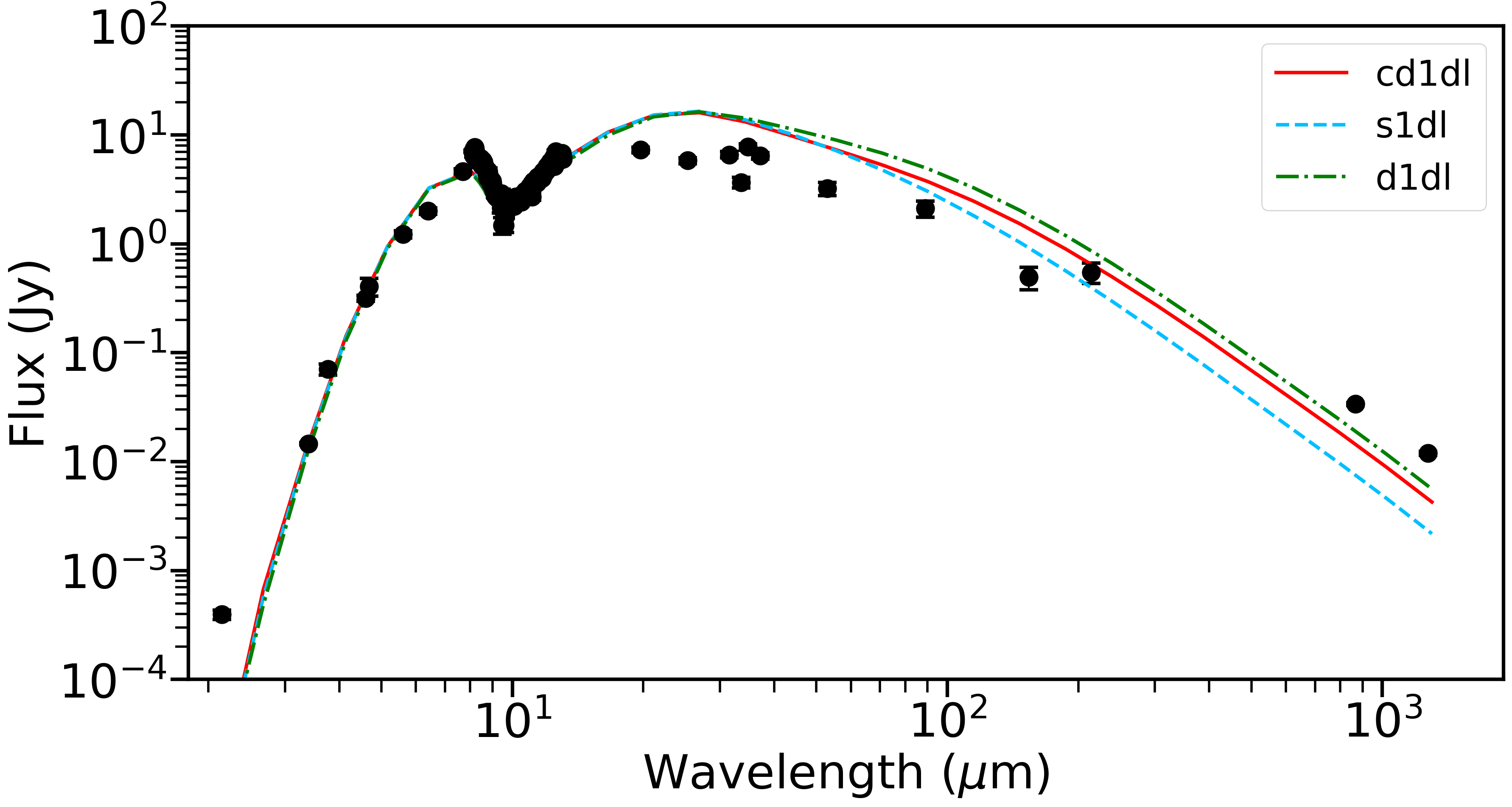}
    \vspace{0.2mm}
    \includegraphics[scale=0.2]{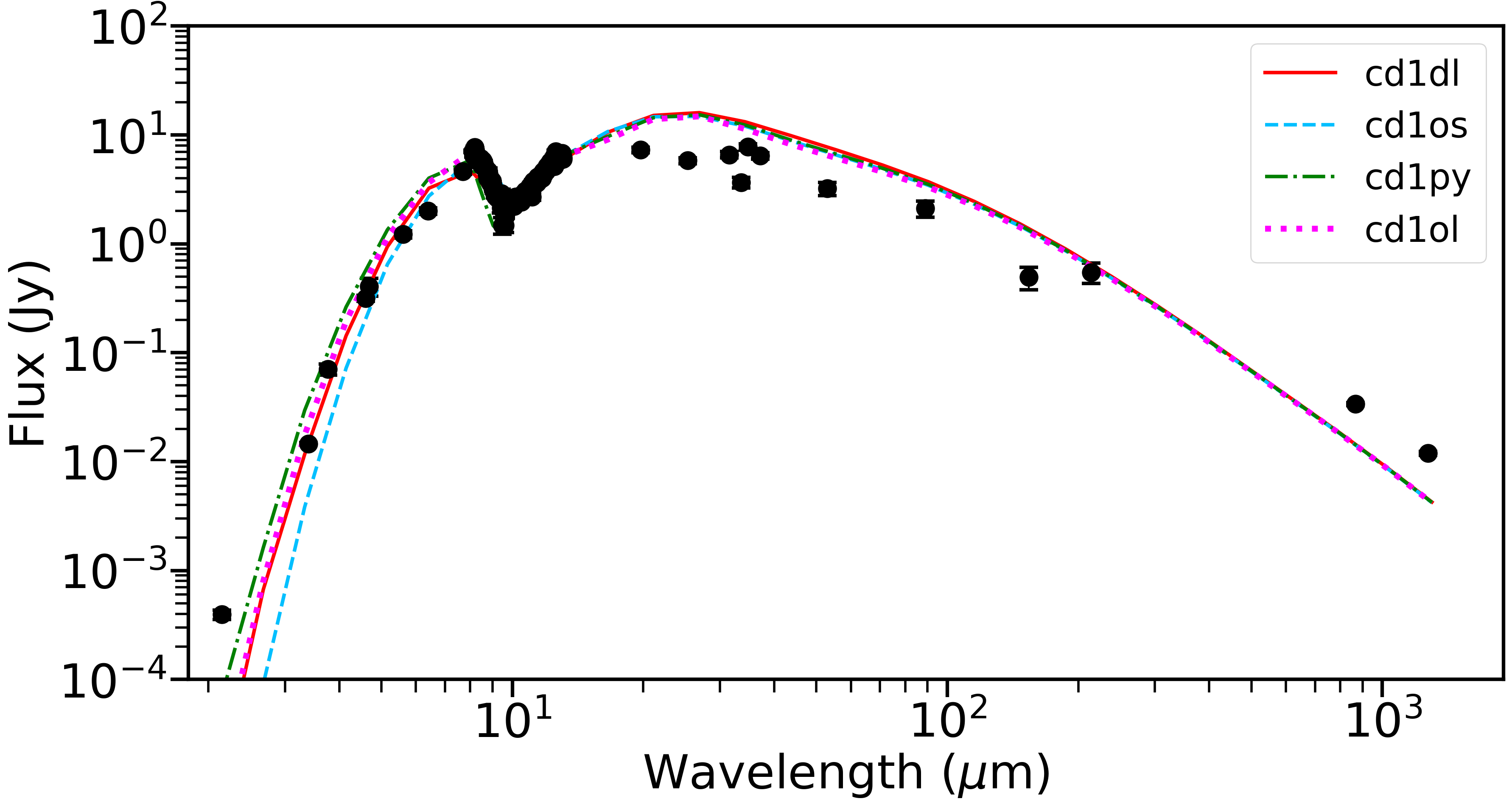}
    \vspace{0.2mm}
    \includegraphics[scale=0.2]{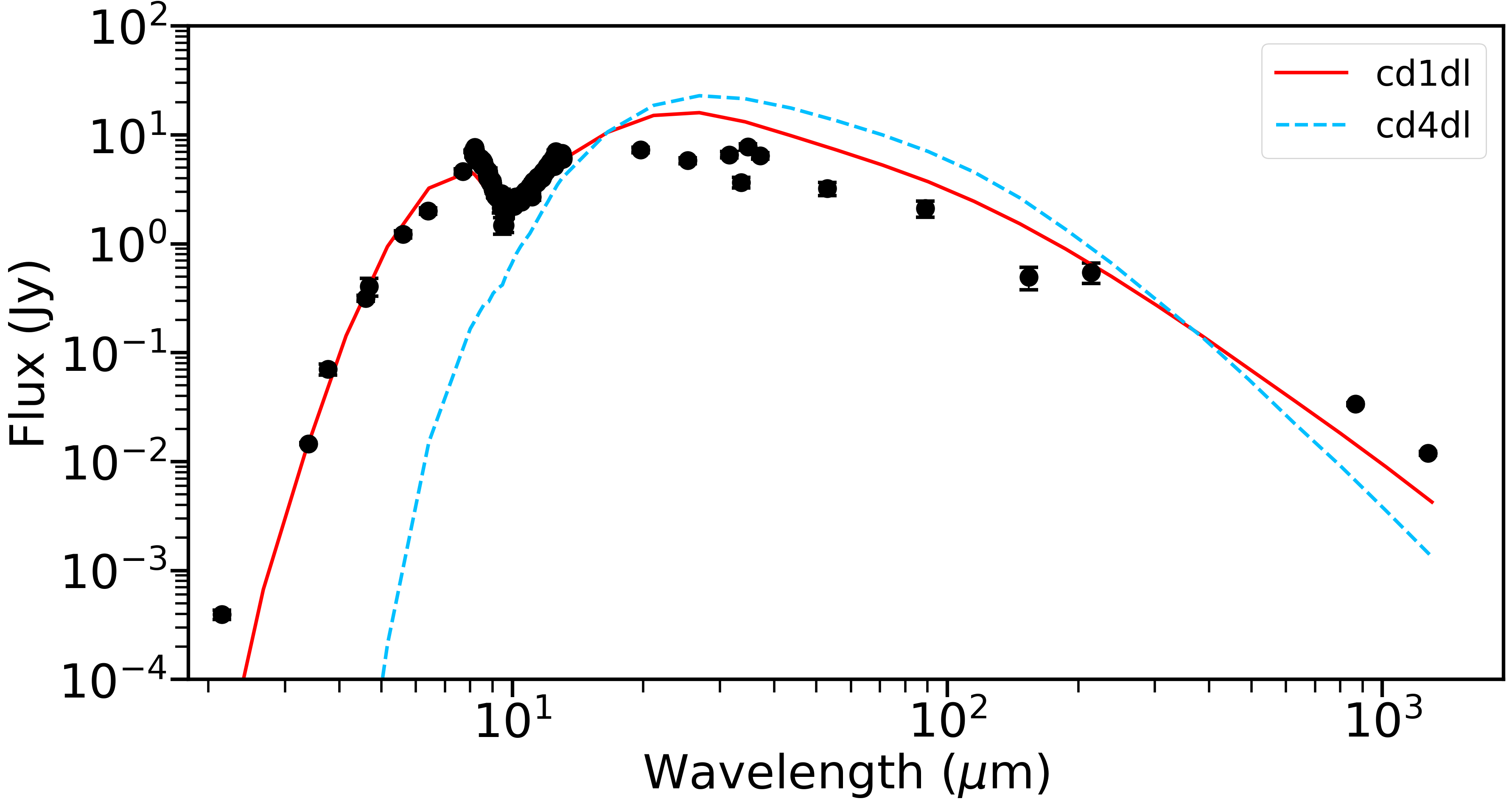}
    \caption{Final SED fits to BLG-360 using RADMC-3D. Each plot compares different models. \textbf{Top-left}: Comparing RADMC-3D models with varying stellar temperature and outer shell density. \textbf{Top-right}: Comparing the different geometries explored. \textbf{Bottom-left}: Comparing the different silicate grain types. \textbf{Bottom-right}: Comparing the inclusion of metallic iron grains to replacing them with carbon grains.}
    \label{fig: radmc3d seds}
\end{figure*}

We present the best-fitting SEDs in Fig. \ref{fig: radmc3d seds} and the input and output parameters in Tables \ref{tab: radmc3d input} and \ref{tab: radmc3d output}. Also presented are models with similar characteristics but with key parameters changed, to compare the impact of those parameters on our model fit. Our best-fitting models were CD models with an extended outer shell (9500 AU) and a large cavity between 1000 and 8500 AU. The slope of the Rayleigh-Jeans tail is the same for both CD and D models, but the cavity suppresses the IR flux between 53--214 \micron\ (covered by SOFIA-HAWC+) for model \texttt{cd1dl}, compared to model \texttt{d1dl}. The slope of our S models is steeper than we see in our observations at $\lambda$ > 30 \micron (see Fig. \ref{fig: radmc3d seds}, top-right panel). The impact of different geometries is negligible for fluxes $\lesssim$ 30 \micron.

\begin{figure}
    \centering
    \includegraphics[trim=0 50 0 0, clip=True,width=\columnwidth]{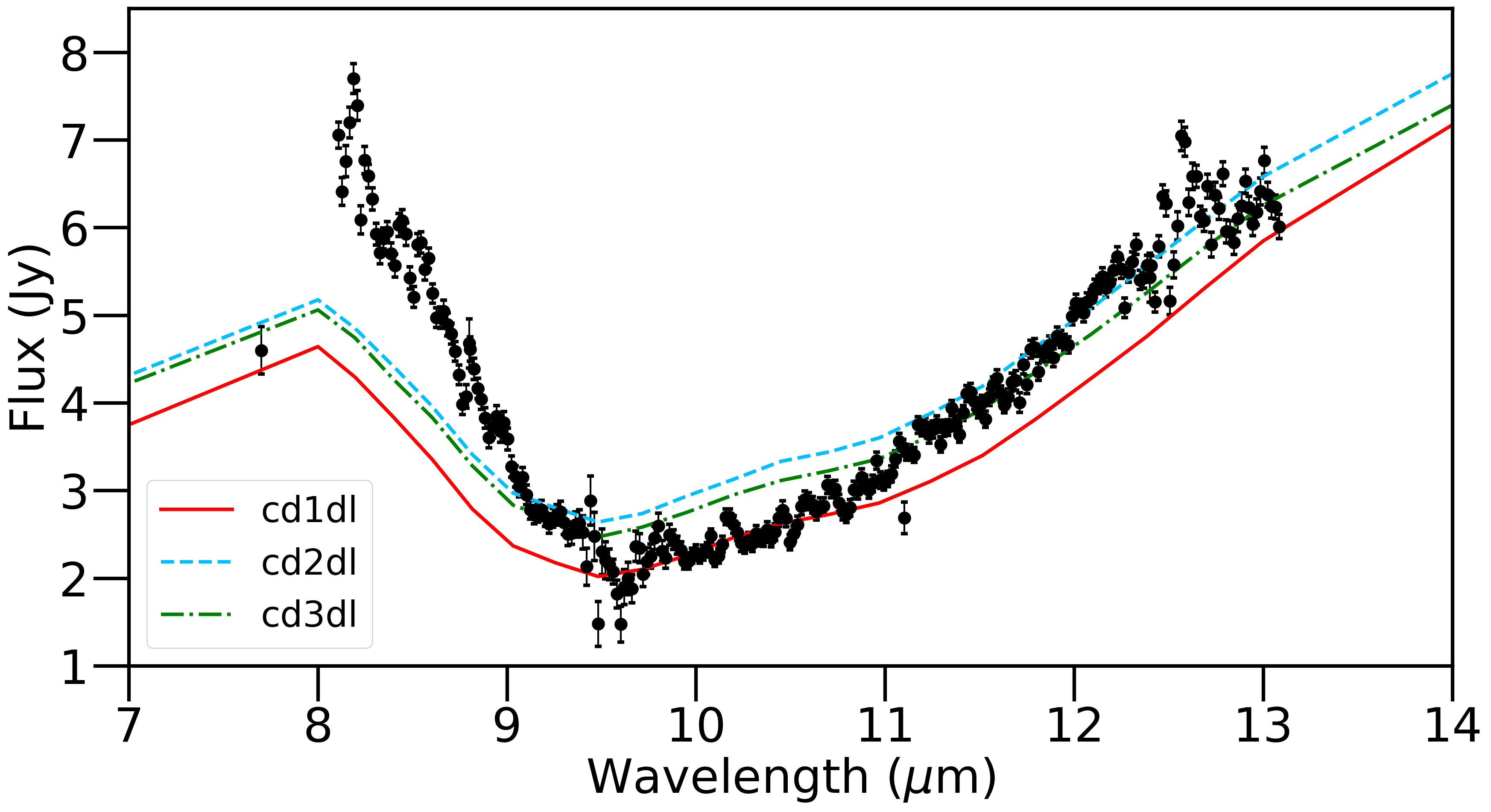}
    \includegraphics[trim=0 50 0 0, clip=True,width=\columnwidth]{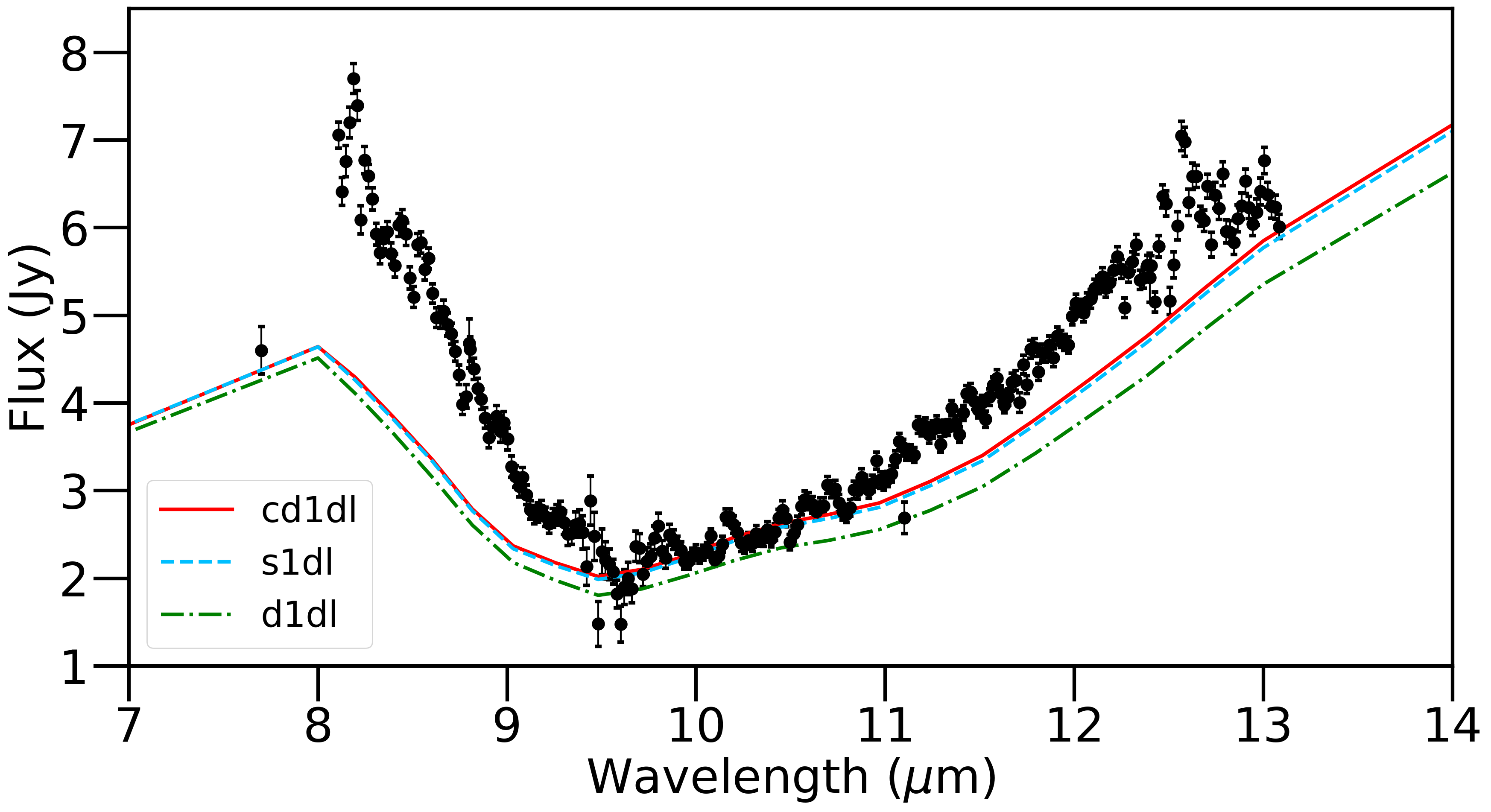}
    \includegraphics[trim=0 0 0 0, clip=True,width=\columnwidth]{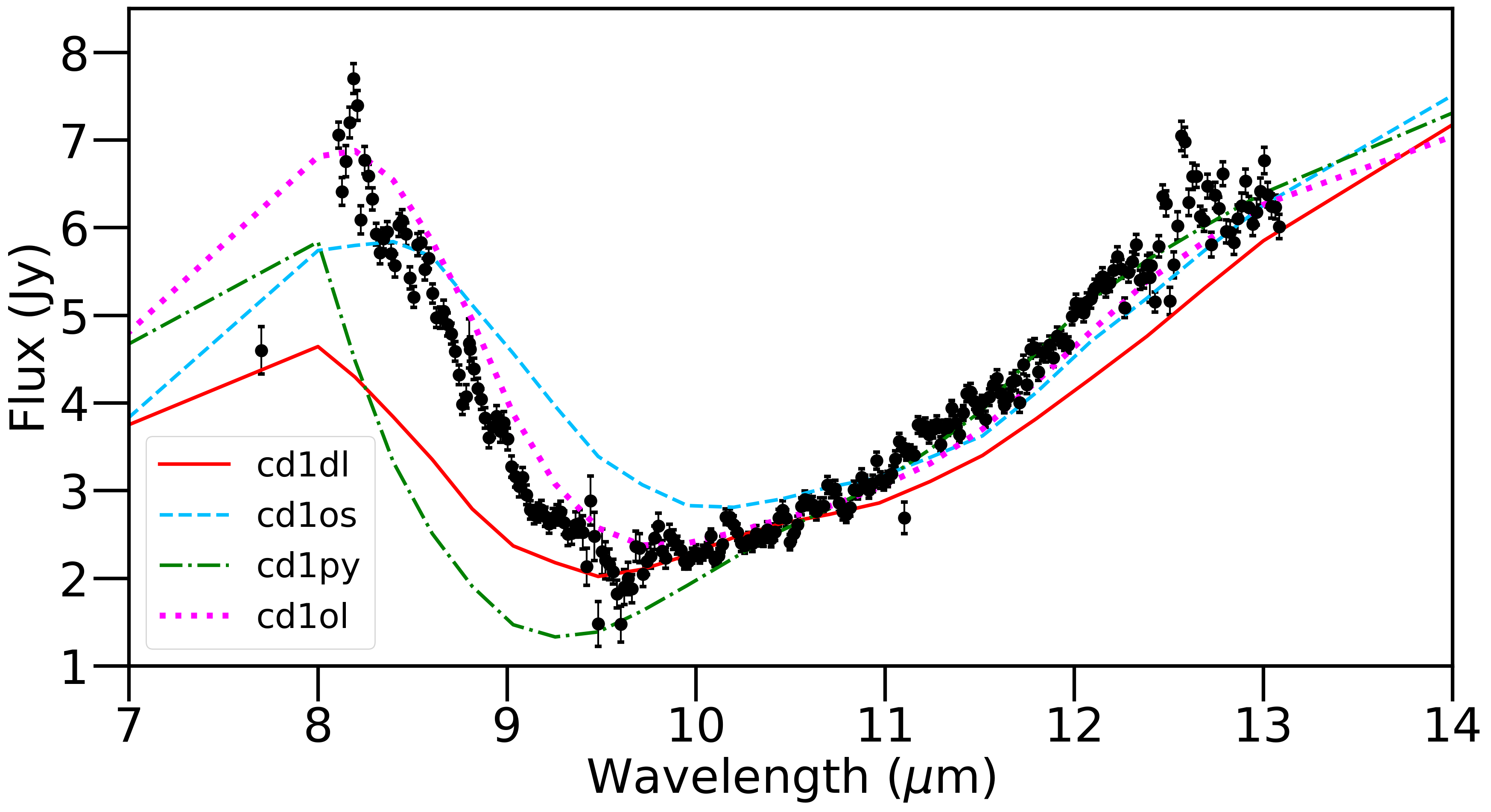}
    \caption{Same as Fig. \ref{fig: radmc3d seds}, but showing in more detail how the models fit to the 10 \micron\ silicate absorption feature. The equivalent plot comparing the iron and carbon models is absent, as the carbon model ({\tt cd4dl}) does not show any silicate absorption.} 
    \label{fig: radmc3d seds 10 micron}
\end{figure}

\begin{figure*}
    \includegraphics[trim={0 30 100 80},clip=True,width=\textwidth]{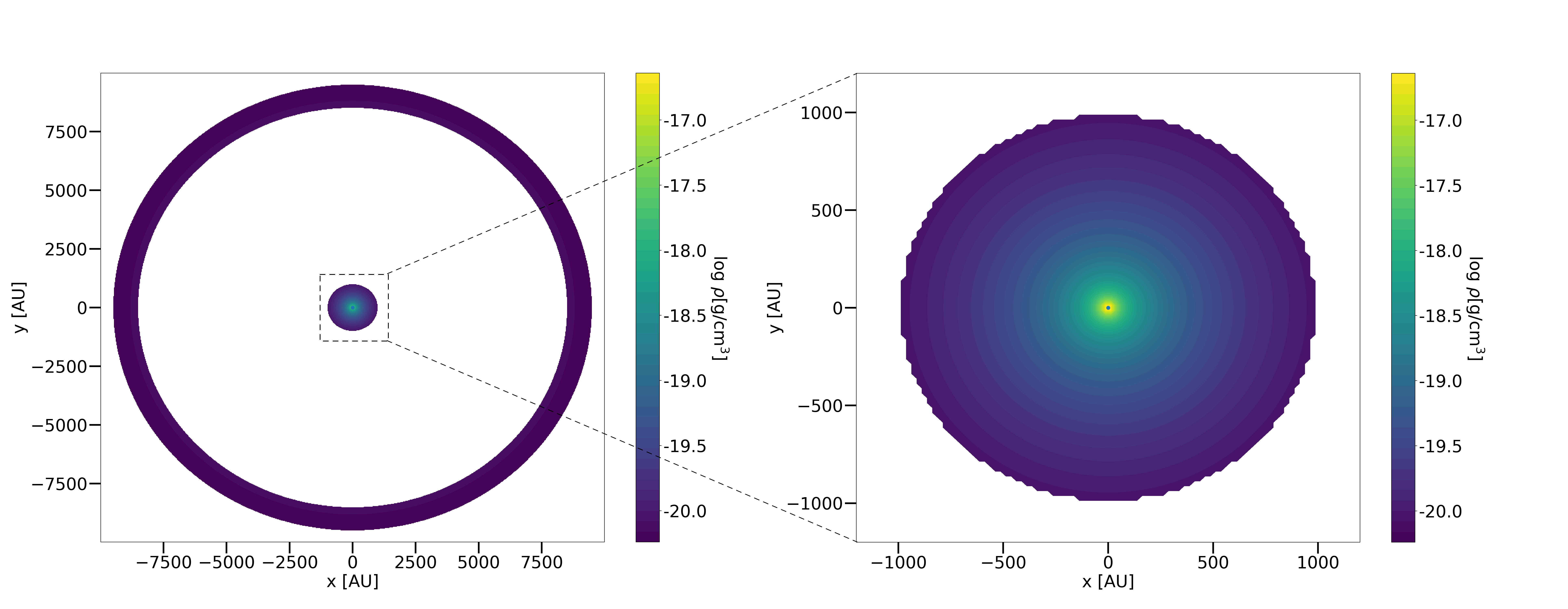}\\
    \includegraphics[trim={0 30 100 80},clip=True,width=\textwidth]{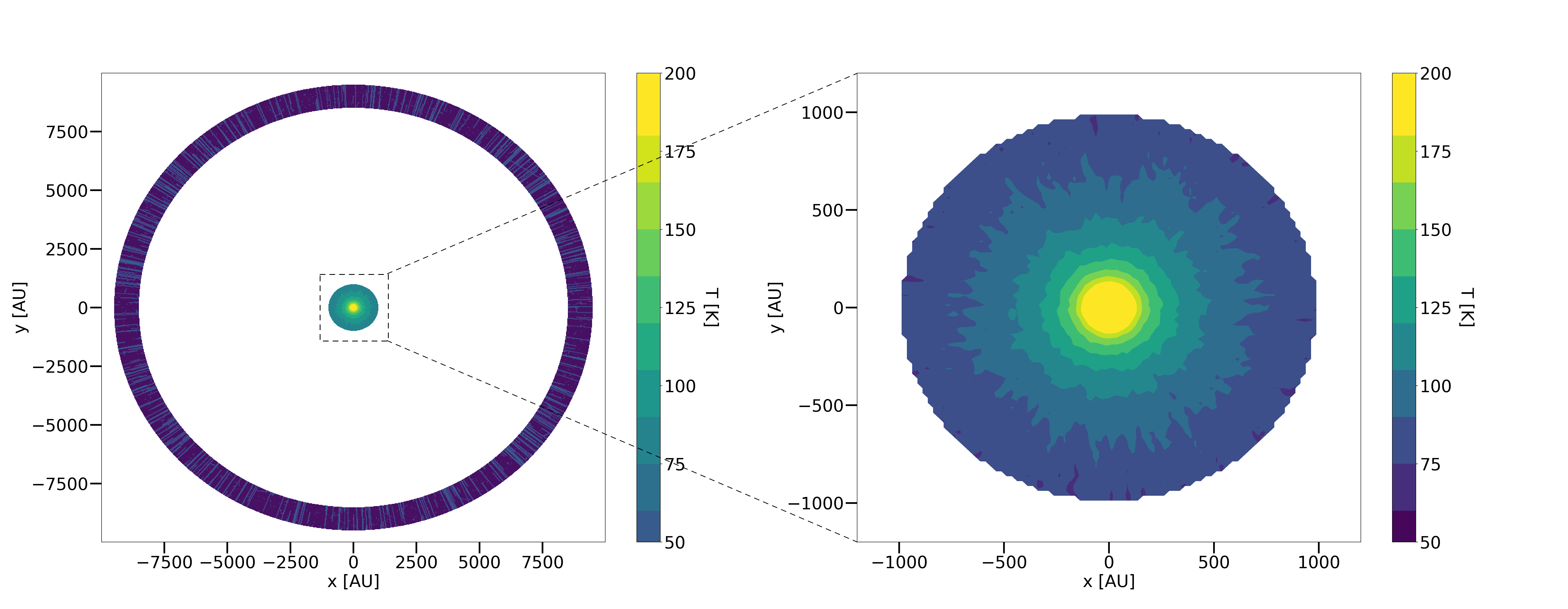}\\
    \caption{Plots indicating the physical structure of model {\tt cd1ol}. The top and bottom row show the dust density and temperature, respectively, whilst the left and right columns show the structure across the full extent of the model, and zoomed in to the inner shell only, respectively. The concentric rays are the result of imperfect sampling with the Monte Carlo methods.}
    \label{fig: cd1ol plots}
\end{figure*}

We find that the best fitting model is {\tt cd1ol}, which is able to match the 10 \micron\ silicate feature as well as the slope in the NIR (see Figs. \ref{fig: radmc3d seds} and \ref{fig: radmc3d seds 10 micron}). Figure \ref{fig: cd1ol plots} shows the density and temperature structure of the model. We see that the dust temperature outside the inner shell does not exceed $\sim$90 K, and only exceeds 200 K at the innermost region (r $<$ 250 AU). The absence of warm (T $>$ 100 K) dust at the outermost regions explains the significant mid-IR, far-IR and sub-millimetre (sub-mm) emission in the SED. The maximum density $\rho_{\rm max}$ is 1.91\standform{-17} g cm$^{-3}$ for all models and decreasing to $\approx$10$^{-19}$--10$^{-20}$ g cm$^{-3}$ at the outer edge of the inner shell. Even with f$_p$=60, the outer shell density is still an average of $\approx$10$^{-20}$ g cm$^{-3}$. The total dust mass across both shells is 0.012 M\solar, with just 2$\times$10$^{-4}$ M\solar\ found within the inner shell, and the rest in the outer shell. 

To constrain the distance, we compare the observed total flux, $F$, to the luminosity of the black body in our model:
\begin{equation}
    L=4\pi\sigma_{\rm SB} R_{\star}^2T_{\rm eff}^4=4\pi d^2 F,
    \label{eq: luminosity}
\end{equation}
where $\sigma_{\rm SB}$ is the Stefan-Boltzmann constant and $d$ is the distance to the star. We then rearrange Equation \ref{eq: luminosity} for $d$ to calculate the distance. For model {\tt cd1ol}, we get $d$=4.09 kpc, which corresponds to a luminosity of 1646 L\solar. Both quantities are lower than in T13.

Our models slightly overpredict the FIR fluxes at 20--220 \micron\ and do not reproduce the flux variations seen in data gathered with SOFIA (e.g., small dips at around 25 and 154 \micron). It was very difficult, if not impossible, to find better-fitting models within our simplifying assumptions. The models are, however, consistent with the observed fluxes within 5$\sigma$ uncertainties. It should be pointed out that modelling such a widely-covered and well-sampled SED of a red novae remnant has not been attempted before and would even be challenging for more standard sources, such as AGB stars.
\renewcommand{\arraystretch}{1.5} 
\begin{table*}
    \caption{Input parameters for RADMC-3D models.}
    \centering
    \begin{tabular}{ccccccccccccccc}\hline
    Model & R$_{\star}$ & T$_{\rm eff}$ & Silicate & Sil:Al$_2$O$_3$:mFe:amC & r$_{in}$ & r$_{out}$ & r$_{in}^{cav}$ & r$^{cav}_{out}$ & r$_{bound}$ & f$_{\rho}$ \\
    & (R\solar) & (K) & type$^a$ & & (AU) & (AU) & (AU) & (AU) & (AU) & \\\hline
    cd1dl & 150 & 3000 & dl-sil & 3:2:7:0 & 10 & 9500 & 1000 & 8500 & 8500 & 60 \\
    cd2dl & 150 & 3500 & dl-sil & 3:2:7:0 & 10 & 9500 & 1000 & 8500 & 8500 & 60 \\
    cd3dl & 150 & 3500 & dl-sil & 3:2:7:0 & 10 & 9500 & 1000 & 8500 & 8500 & 100 \\
    cd4dl & 150 & 3000 & dl-sil & 3:2:0:7 & 10 & 9500 & 1000 & 8500 & 8500 & 60 \\
    cd1os & 150 & 3000 & oss-sil & 3:2:7:0 & 10 & 9500 & 1000 & 8500 & 8500 & 60 \\
    cd1py & 150 & 3000 & pyroxene & 3:2:7:0 & 10 & 9500 & 1000 & 8500 & 8500 & 60 \\
    cd1ol & 150 & 3000$^c$ & olivine & 3:2:7:0 & 10 & 9500 & 1000 & 8500 & 8500 & 60 \\
    s1dl & 150 & 3000 & dl-sil & 3:2:7:0 & 10 & 9500 & - & - & - & 1$^b$ \\
    d1dl & 150 & 3000 & dl-sil & 3:2:7:0 & 10 & 9500 & - & - & 8500 & 60 \\\hline
    \end{tabular}
    \tablefoot{
    $^a$ References for different silicate grains are: dl-sil: \citet{drainelee1984}, oss-sil: \citet{ossenkopf1992}, pyroxene: \citet{dorschner1995olivinepyroxene}, olivine: \citet{dorschner1995olivinepyroxene}. $^b$ f$_{\rho}$=1 is implemented to remove the outer shell. $^c$ Reproducing the cd1ol model with a synthetic stellar spectrum indicates that the central star is more likely to have a temperature of 3500 K. See Sect. \ref{sect: T13 results discussion}.}
    \label{tab: radmc3d input}
\end{table*}
\begin{table}[]
    \centering
    \caption{Output properties of our RADMC-3D models.}
    \begin{tabular}{cccc}
\hline
Model & Dust mass & Luminosity & Distance \\
& (M\solar) & (L\solar) & (kpc) \\\hline
cd1dl & 0.012 & 1646 & 4.09 \\
cd2dl & 0.012 & 3050 & 5.57 \\
cd3dl & 0.019 & 3050 & 5.57 \\
cd4dl & 0.012 & 1646 & 4.09 \\
cd1os & 0.012 & 1646 & 4.09 \\
cd1py & 0.012 & 1646 & 4.09 \\
{\bf cd1ol} & {\bf 0.012} & {\bf 1646} & {\bf 4.09} \\
s1dl & 0.0018 & 1646 & 4.09 \\
d1dl & 0.013 & 1646 & 4.09 \\\hline
    \end{tabular}
    \label{tab: radmc3d output}
\end{table}
\section{Discussion}\label{sect: discussion}
\subsection{Caveats}\label{sect: caveats}
Before discussing the implications of our model, it is important to address the shortcomings. Although we are able to make some constraints on the stellar properties of the merger remnant, these constraints carry some intrinsic uncertainty, both from the observational errors and the methodology.

The top two panels of Fig. \ref{fig: radmc3d seds} show that differing stellar properties and dust distributions can satisfactorily reproduce most of the observed SED at $\lambda <$ 215 \micron\ within 5$\sigma$. 
At NIR wavelengths, the difference between the {\tt cd1dl}, {\tt cd2dl}, and {\tt cd3dl} models are minor, with {\tt cd1dl} providing a slightly better fit to the slope we see for $\lambda <$ 8 \micron. Likewise, the comparison of the different dust distributions shows no difference to the SED at $\lambda \lesssim$ 30 \micron. Only the slope of the Rayleigh-Jeans tail is somewhat sensitive to the dust distributions considered.

We defined the wind density structure to resemble that of a spherical wind, meaning that the density $\rho$ scaled with radial distance $r$ as $\rho \propto r^{-2}$. We also arbitrarily assume that the dust sublimation temperature T$_{sub}$=1700 K. In reality, dust above this temperature cannot form, leading to a spherical cavity around the star in which no dust can form. At T$_{sub}$=1700 K, we do not see any inner cavity. We therefore introduce an inner cavity manually, defined by the dust inner radius r$_{in}$, which is chosen only to provide better fitting to the model, and is not constrained by the choice of T$_{sub}$. To obtain a more reliable constraint, we would need a better spatial resolution, which would require higher computing power to run our models.
\subsection{Spherical vs disk geometry} \label{sect: sphere v disk}
In order to limit the number of free parameters, we assumed in our RADMC-3D models a spherical geometry for the dust shells. However, envelopes formed by interacting binaries may, in principle, display a wide range of geometries.

Stellar mergers require the loss of angular momentum from the system in order to produce the inspiral phase during the common-envelope (CE) evolution. This often takes place via mass loss through the second Lagrange point (L$_2$), which forms a spiral wind in the orbital plane \citep{nandez2014,pejcha2014,macleod2022}. The material ejected in this wind can then later condense into a dusty torus as the stellar remnant cools. Several red nova remnants have shown evidence of a toroidal structure, such as V4332 Sgr \citep{kaminski2010v4332} and V838 Mon \citep{kaminski2021v838mon}. \citet{pejcha2017} also showed that such spiral winds could explain V1309 Sco's pre-merger light curve.

However, recent simulations of the subsequent evolution of the ejected material after the inspiral phase have shown how the torus can later evolve into a spherical structure. Analytical models by \citet{glanz2018} show that the CE can expand through slow, steady mass loss driven by dusty winds rather than being dynamically ejected, and that this would have similar appearances to the circumstellar envelopes of AGB stars (i.e., spherical). Hydrodynamic simulations of two pre-merger CE systems, which differ by the masses of the AGB in the binary system, indicate a difference in the morphology of the CE depending on whether the expansion is dust-driven or driven through recombination energy \citep{bermudezbustamante2024}. In contrast to \citet{glanz2018}, \citet{bermudezbustamante2024} find that dust-driven expansion remains elongated in the equatorial plane, whereas recombination energy drives the expansion into a spherical shell, even showing a clear cavity between the dust shell and the central binary. However, the \citet{bermudezbustamante2024} models do not show any inner shell. This presents the possible scenario that the inner and outer shells in BLG-360 have differing origins, and that the inner shell is more likely to be formed from either merger ejecta or mass loss via winds from the merger remnant. The size of the inner shell is consistent with this scenario, as we discuss in Sect. \ref{sect: dust extent discussion}. The dust mass yields of the \citeauthor{bermudezbustamante2024} models are of the order of 10$^{-2}$ M\solar, comparable to our results and therefore supporting the evolution scenario presented by \citet{bermudezbustamante2024}.

Using a smoothed particle hydrodynamics (SPH) code, \citet{iaconi2019} examined the evolution of material lost during the inspiral phase. They showed that initially, the material lost during inspiral forms a spiral pattern \citep[see also][]{nandez2014, pejcha2014}, but after $\approx$9000 days the spiral pattern disappears and forms a spherical distribution. \citet{iaconi2020} examined the dust formation in the same simulations as \citet{iaconi2019}, and found that the majority of dust is formed between 300 and 5000 days, meaning that the dust is formed before the distribution of circumstellar matter becomes spherical. Their grain sizes are an order of magnitude smaller than what we find in our RADMC-3D model, as is the dust mass. The SPH simulations were run up to $\approx$50 years. It is possible that the inspiral phase could have lasted longer for BLG-360 than the timescales covered in the \citeauthor{bermudezbustamante2024} and \citeauthor{iaconi2019} simulations, and that in the subsequent 22 years since the merger event there was additional dust formation from merger and post-merger outflows (see Sect. \ref{sect: T13 results discussion}).

In order to rule out a disk-like dust structure for BLG-360, we attempted to fit the observed SED with the SED Fitter python-based fitting tool\footnote{https://sedfitter.readthedocs.io/en/stable/}. We compared SED models calculated using the Hyperion code \citep{robitaille2011hyperion} and different dust geometries. A full description of the models is given in \citet{robitaille2017}. We searched for the best match to our observed SED among the model sets designated as sp-hmi and spu-hmi. The sp-hmi models feature a disk with a cavity between the inner radius and the central source, with the inner radius varied as a free parameter. The spu-hmi models feature the same disk structure, along with a rotationally flattened envelope as defined by \citet{ulrich1976}.

\begin{figure}
    \centering
    \includegraphics[trim=0 0 0 0, clip=True, width=\columnwidth]{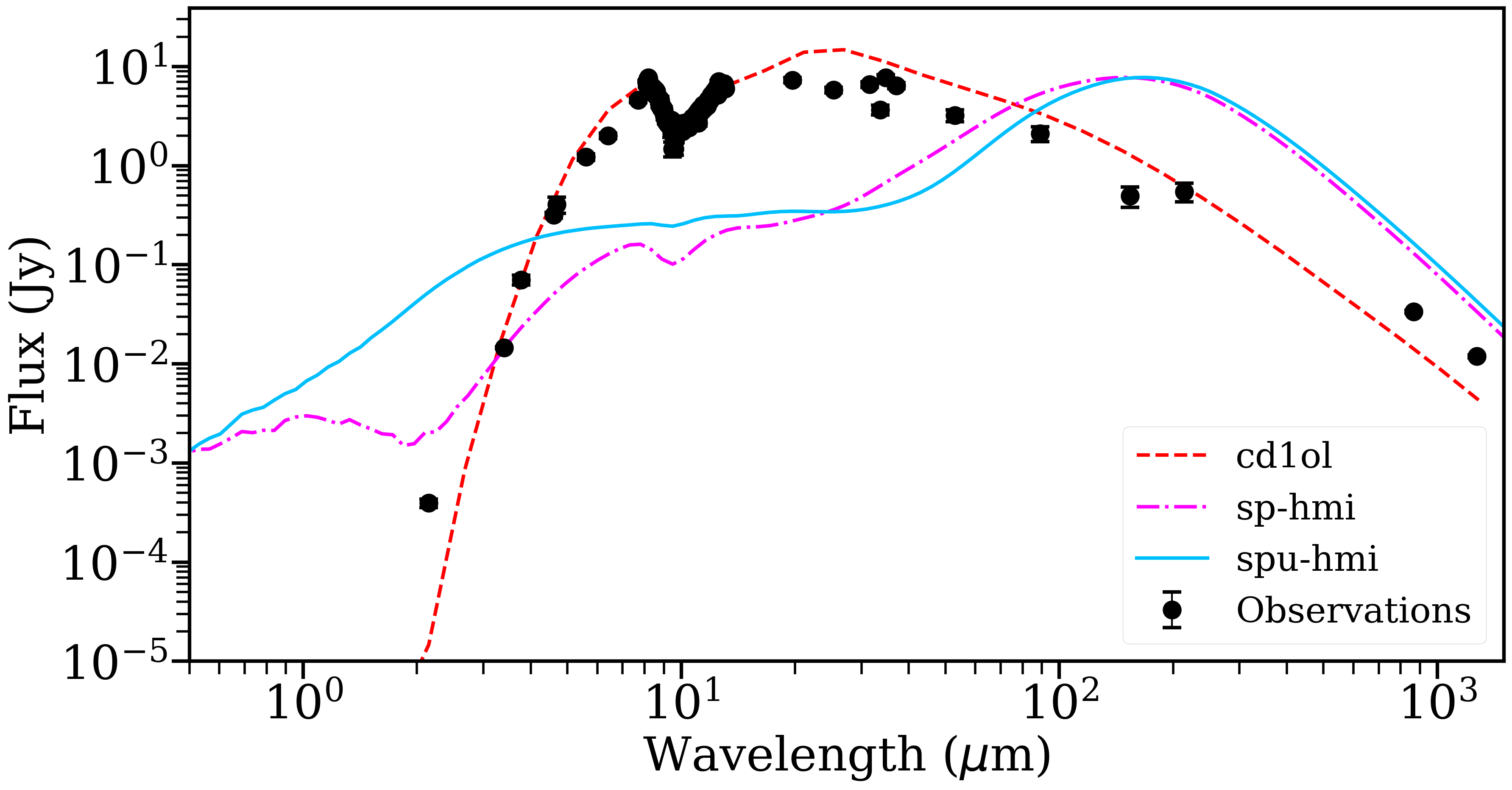}
    \caption{Hyperion best models and our RADMC-3D best model ({\tt cd1ol}) overlaid on the observed SED. In magenta and blue, respectively, we plot the best-fitting Hyperion models using SED Fitter for the sp-hmi and spu-hmi model sets. The Hyperion models are normalised  and scaled up to the maximum flux in the observed SED.} 
    \label{fig: radmc3d + sedfitter}
\end{figure}

We compare the best fitting disk models from Hyperion to our observations and the best RADMC-3D spherical model {\tt cd1ol} in Fig. \ref{fig: radmc3d + sedfitter}. The observations and model {\tt cd1ol} are plotted in absolute flux units, as in Fig. \ref{fig: radmc3d seds}, whereas the Hyperion models are normalised and scaled to the maximum observed flux, so the shapes of the SEDs can be compared. In the Hyperion models, we see a much larger far-IR and sub-mm excess than we see in RADMC-3D relative to the 10 \micron\ absorption feature, whilst significant NIR and optical emission are also present. The significantly higher optical emission is due to dust scattering above and below the plane of the disk. The 10 \micron\ absorption feature is also weaker, especially for the spu-hmi model. The shape of the SED, coupled with the fact that BLG-360 is not visible at optical wavelengths, immediately rules out a disk geometry dominating the dust structure of BLG-360. We cannot rule out some form of disk structure within the inner shell in our RADMC-3D model, as such a structure may be hidden by the combined optical depth of both shells. However, in such a scenario, we would not expect the presence of a compact disk to significantly affect the shape of the SED.

\subsection{Dust extent}\label{sect: dust extent discussion}
One of the main advantages of the RADMC-3D models is that we can explain the SED using dust extending out to a maximum distance of 9500 AU, rather than 20 pc that we found using DUSTY (Appendix \ref{appendix: dusty}). At a distance of 4.09 kpc, a radius of 9500 AU would subtend an angle of $\approx$2\farcs3, whilst for a distance of 5.57 kpc (as for models {\tt cd2dl} and {\tt cd3dl}), 9500 AU is equivalent to 1\farcs7. A source with this size would be unresolved by the SMA and ACA beams. We find it unlikely that the outer shell is much smaller than in the presented models. Indeed, if we implemented a more compact outer shell, the slope of the Rayleigh-Jeans tail would be steeper and so the sub-mm fluxes, as well as the SOFIA-HAWC+ fluxes, would be underestimated.

The outer radius of the inner shell r$_{in}^{cav}$=1000 AU, is consistent with the radius of a dust shell produced by post-merger mass loss, based on previous observations of other red novae. A larger inner shell suppresses the NIR fluxes, and vice versa. Based on the time between the eruption and the most recent observations in which the source was detected (NEOWISE 4.6 \micron; 23/09/2023) of 7654 days, as well as the average outflow velocity observed in red novae \citep[200 km s$^{-1}$; ][]{kaminski2018}, we would expect dust carried outwards via such outflows to be 884 AU from the star. For an outer radius of 1000 AU, the average velocity would have to be 226 km s$^{-1}$, consistent with expectations for the outflows observed within post-merger environments of non-compact stars.

For the outer shell, if we assume that the material was ejected during the merger, the average velocity would be $\approx$2150 km s$^{-1}$, which is fast even for red novae such as CK Vul \citep{kaminski2021ckvul,tylenda2024}, although still not of the order of SNe \citep[$\sim$3000-15000 km s$^{-1}$;][]{mazzali2005,smartt2009}. It is more likely that the material in this shell originates from pre-merger mass loss, as we describe in Sect. \ref{sect: sphere v disk}, especially since T13 advocated a red giant progenitor for BLG-360. If we assume the velocity is of the same order as that expected for red novae (200 km s$^{-1}$), the time taken to reach 9500 AU would have been $\approx$225 yrs. Therefore, material that formed the outer shell must have been lost from the star for many years prior to the merger event, as 225 yr is inconsistent with the timescale for the inspiral phase in CE systems based on observations and modelling of the progenitor of V1309 Sco \citep{tylenda2011,pejcha2014}, as well as the exponential period decline that signified the inspiral phase spanning 6 years, as well as modelling of the inspiral phase \citep{iaconi2019,bermudezbustamante2024}. A normal wind from a red giant progenitor would be an order of magnitude slower, and therefore the outer shell would require mass loss that occurred thousands of years ago. Such a relic envelope was postulated to exist around the red nova remnant V1309 Sco \citep{tylenda2016,masonshore2022}, whose progenitor was also a red giant \citep{stepien2011}. The dusty outer shell was revealed owing to FIR observations with the Herschel telescope. If the relic envelope is a common feature of red nova progenitors, it could have been easily missed if the object was not observed at FIR and longer wavelengths.

\subsection{Mineralogy: silicate dust}\label{sect: silicates discussion}
The deep absorption detected at $\approx$10 \micron\ is a typical feature of circumstellar environments dominated by silicate dust, and is characteristic of thick dust shells \citep{suh1999}. Using the silicate classification of \citet{egansloan2001}, the feature in BLG-360 is `classic'. However, the exact nature of the silicate dust, including its optical properties, is often difficult to constrain. It has been found that different opacity tables are necessary for dust with different temperatures \citep[in the case of spherical dust shells, also for dust at different distances from the central star; see][]{suh1999}. However, for simplicity, we use the same opacity curves for all dust within each model.

We see how the model SED changes for different silicate types (astro-silicates, Ossenkopf, pyroxene, and olivine) in the bottom-left panel in Fig. \ref{fig: radmc3d seds} and the bottom panel in Fig. \ref{fig: radmc3d seds 10 micron}. The former shows very little difference in the overall shape in the SED, with some minor change in the slope of the SED in the NIR, whilst almost no change at all is seen at longer wavelengths. The latter, however, shows how the shape of the 10 \micron\ absorption feature changes. One common feature in the majority of models is that the blue wing of the 10 \micron\ feature is often underestimated, as we see for pyroxene and the \citet{drainelee1984} `astro-silicates' models. The \citet{ossenkopf1992} silicates tables better fit the blue wing, but are not able to fit the rest of the absorption profile, including the centroid. The only silicate type that traces the absorption profile to a high degree of accuracy is amorphous olivine \citep{dorschner1995olivinepyroxene}. The modelling of amorphous silicates is consistent with the low-temperature environment (50--200 K, see Fig. \ref{fig: cd1ol plots}), as high temperatures can lead to either crystalline silicate formation or the conversion of amorphous silicate dust to an ordered, crystalline structure \citep{fabian2001}.

Although significant silicate dust is necessary to produce the 10 \micron\ absorption feature, we also require significant alumina dust to reproduce the absorption profile. This was also found from our DUSTY SED models (see Appendix \ref{appendix: dusty}). We find that a silicate to alumina ratio of 3:2 is able to reproduce the depth and width of the 10 \micron\ profile (see Fig. \ref{fig: radmc3d seds 10 micron}), whereas a pure silicate feature would be too narrow. Alumina dust may be common in red nova remnants \citep[e.g.,][]{banerjee2015}.

\subsection{Mineralogy: iron vs carbon}\label{sect: iron v carbon discussion}
The majority of our models contain dust composed of olivine silicates, alumina, and metallic iron in the ratios 3:2:7, and are thus dominated by the metallic iron component (except for {\tt cd4dl}). Here we discuss this high content of iron grains.

It is clear from the reduced NIR emission and lack of optical emission that the dust has high opacities at wavelengths $\lesssim$ 8 \micron. A common source of high NIR opacities in astrophysical environments is carbon, commonly seen around carbon-rich AGB stars \citep{tosi2023}. BLG-360, like most other red nova remnants, is oxygen-rich, and so it would be unusual to have a high carbon content locked in dust. We instead investigated metallic Fe grains (mFe) as the source of NIR opacities, which fitted well but required a high abundance (58\%). To ensure that there is no degeneracy between iron and carbon models, we investigated the impact of including carbon as the source of high NIR opacities instead of metallic iron. In Fig. \ref{fig: radmc3d seds} (bottom-right panel), we compare models {\tt cd1dl} and {\tt cd4dl}. The only difference is that we exchange the same quantity of iron dust for amorphous carbon dust (see Table \ref{tab: radmc3d input}). The result is that our opacities are too high for wavelengths $\lesssim$17 \micron, resulting in underestimated NIR fluxes and no 10 \micron\ feature. We also experimented with including more silicate grains with a lower composition of carbon grains, but this still did not replicate the NIR fluxes. The result of including a higher composition of silicates was a narrower 10 \micron\ absorption feature, as well as the emergence of a new absorption feature at $\approx$ 18 \micron, which we do not see in our observations. Therefore, we conclude that our source of NIR opacities is more likely to be solid iron than carbon grains.

Iron dust has been identified in the vicinity of evolved stars \citep[e.g.,][]{mcdonald2010,speck2015}, and iron dust formation models have been formalised for composite grains \citep{gail1999} and solid iron \citep{verhoelst2009}. \citet{kemper2002} propose that it is possible that iron dust can condense in the winds of solar-type stars, based on the abundances of iron within meteorites. As metallic iron and silicates have similar condensation temperatures ($\approx$50--100 K difference), they propose that silicates and metallic iron can form simultaneously. This is used to explain 4\% Fe abundance in the dusty CSE around the OH/IR star OH 127.8+0.0, but could not explain the dominating abundance of metallic iron in BLG-360 that we observe. \citet{speck2015} find that a silicate:iron mix of 3:1 can represent the inner dust around the AGB star HD 161796, as well as a 2-shell geometry, albeit without a cavity between the two shells. DUSTY modelling of the O-rich AGB star RT Vir also revealed a 2-shell geometry with a significant amount of iron dust \citep{preston2024}, and interestingly, an outer shell at distances of the order of $\approx$10$^4$ AU, similar to BLG-360. Analysis of pulsating stars in globular clusters \citep{mcdonald2010} show that many such stars exhibit a mid-IR excess that is likely due to solid iron dust. A study of M-type giants by \citet{marini2019} in the Large Magellanic Cloud (LMC) found that many of the stars exhibited unusual SEDs that could not be explained by silicate-dominated dust. They suggest that, through third dredge-up \citep[e.g.,][]{uttenthaler2024} and hot bottom burning \citep[HBB;][]{ventura2011}, higher mass AGB stars ($>$ 4 M\solar) have depleted oxygen and magnesium abundances at the surface. This inhibits silicate formation, and destroys surface carbon in the star, leaving HBB products such as iron as the dominant dust condensation species. Iron dust has also been suggested to be an important constituent in envelopes of even more massive stars, such as red supergiants \citep[e.g., in VY CMa;][]{Harwit}, and in supernova remnants \citep[e.g.,][]{dwek2004}.

Considering the mechanism proposed by \citet{marini2019}, the dominance of solid iron dust can be explained by a progenitor primary that has experienced third dredge-up and HBB, with the presence of silicates explained by an oxygen-rich secondary star with a lower mass or experiencing an earlier evolutionary stage. This would point towards the primary being an AGB star. However, SED modelling of the progenitor by T13 indicated that the progenitor luminosity was 290 L\solar\ at a distance of 8.2 kpc (and even four times less at our preferred distance of 4.09 kpc), inconsistent with an AGB star where the expected luminosity would exceed 1000 L\solar. It is also quite likely that the opacities that are currently available of metallic iron do not reproduce the solids in BLG-360 at an adequate level. Indeed, obtaining realistic dust opacities for more standard astronomical sources, like AGB stars and red supergiants, remains one of the biggest sources of uncertainties in modelling of their SEDs and spectra \cite[e.g.,][]{speck2015}.

Overall, we find that the source of our NIR opacities is best represented by iron grains, rather than carbon. The production of solid iron grains is possible from evolved AGB stars that experience third dredge-up and HBB, but the estimated progenitor luminosity is not consistent with an AGB star. It is not clear what the source of iron in this source could be, but could be related to the chemical mixing of the progenitor components during the merger. It could also be that our understanding of the optical properties of astronomical dust is not yet complete.
\subsection{Stellar parameters}
For convenience, the stellar spectrum in our RADMC-3D models was assumed to be Planckian. To test the difference between using a blackbody-approximated star and a more realistic synthetic stellar spectrum, we construct models using synthetic stellar spectra from the BT-Settl (AGSS2009) grid \citep{allard2012}, extracted from the Spanish Virtual Observatory (SVO) service\footnote{http://svo2.cab.inta-csic.es/theory/newov2/index.php?models=bt-settl}. We examine models with log($g$)=0.0, and [X/H]=0.0 (where X is metal content), and look at stellar temperatures between 3000--4500 K, at 500 K intervals. The 3500 K spectrum produces an almost identical SED to the {\tt cd1ol} SED, in which the star was modelled as a 3000 K blackbody. As the spectral features of the synthetic spectra are not seen in the reprocessed SED, it seems that all stellar photons are either scattered or absorbed, and we instead see only re-radiated dust emission. As the synthetic spectrum is a more realistic approximation of the remnant star in BLG-360, we therefore conclude the star has moved back towards an M3 spectral type, compared to the M6 found by T13 for the 2010 epoch, but uncertainties in both estimates are large. The evolution is further discussed in the next section.

\section{Remnant evolution}\label{sect: T13 results discussion}
\begin{figure}
    \centering
    \includegraphics[trim=0 0 0 0, width=\columnwidth]{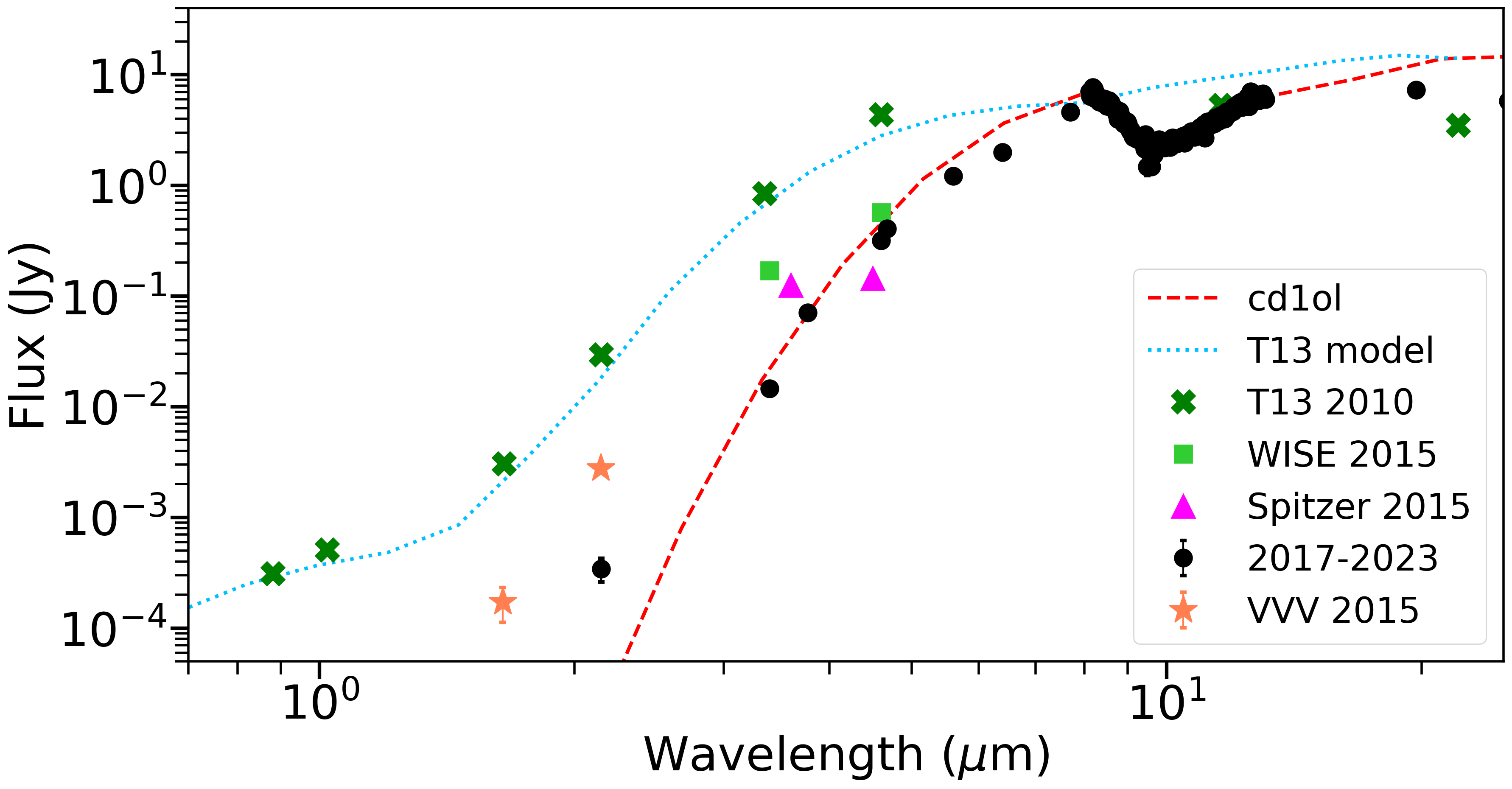}
    \caption{Comparison of data and models for different SED epochs. The 2010 observations (green) and best-fitting 2010 model (cyan) from T13 can be compared to the full observed SED a decade later (black) and our {\tt cd1ol} model (red). We also show WISE and Spitzer fluxes from 2015 in green and magenta squares, respectively. VVV fluxes are shown in orange.} 
    \label{fig: t13 + spitzer + obs}
\end{figure}

To examine the temporal changes of the remnant, we revisit the SED analysis of T13 within our RADMC-3D framework. To make a more direct comparison with our more recent SED, we therefore model the T13 SED from 2010 using RADMC-3D and with our \texttt{cd1ol} model as the starting point. Our best RADMC-3D model for the 2010 SED is shown in Fig. \ref{fig: t13 + spitzer + obs}. We assume that the dust composition and grain size did not change. As the origin of the outer shell is from pre-merger mass loss, we did not change the physical parameters of the outer shell. We therefore changed only the density distribution, extent of the inner shell, and the properties of the star. The key results of this model are shown in Table \ref{tab: t13 radmc3d}.

\begin{table}
    \caption{Input parameters for our best RADMC-3D model representing the 2010 and 2017--2023 epochs.}
    \centering\normalsize
    \begin{tabular}{ccc}\hline
    Parameter & 2010 & 2017--2023\\\hline
    R$_{\star}$ (R\solar) & 150 & 150\\
    T$_{\star}$ (K) & 3200& 3000\\
    r$_{in}$ (AU) & $\leq$ 20 & 10\\
    r$_{in}^{cav}$ (AU) & 700 & 1000\\
    $\rho_{max}$ (g cm$^{-3}$) & 5.72\standform{-18} & 1.91\standform{-17}\\
    Dust mass (M\solar) & 3.44\standform{-3} & 0.012\\
    Stellar luminosity (L\solar) & 2126 & 1646\\\hline
    \end{tabular}
    \label{tab: t13 radmc3d}
\end{table}

Comparing results of both epochs modelled in RADMC-3D (Table \ref{tab: t13 radmc3d}), we find that the central star has not changed in size between 2010 and our 2017--2023 observations. The star may have been slightly hotter (3200 K) in the earlier phase, consistent with the assumed temperature of T13. Additionally, comparing the two epochs in RADMC-3D, we find that the star was $\approx$30\% more luminous in 2010 compared to 2017--2023.

We also find that the star is considerably smaller in both our models in RADMC-3D (150 R\solar\ for both epochs) than found in T13 (300 R\solar), making the star much less luminous than previously assumed (our revised distance lowers the luminosity even further). This discrepancy is of course a consequence of different methodologies.

The inner shell structure has changed between 2010 and 2017--2023. In our 2010 models, we establish an upper limit of 20 AU for r$_{in}$, as larger radii than this significantly increase the flux at sub-\micron\ wavelengths. We found that the inner shell was more compact in 2010, with an outer radius of 700 AU, and also a factor of $\approx$3 less dense than in our 2017--2023 model. Using the WISE observations taken on 19--21 March 2010 as the zero point of the 2010 epoch, we took the first and last observation dates of our 2017--2023 SED and estimated the expansion velocity based on the time difference. Assuming a constant expansion velocity, we estimated the dust expansion velocity to be between 116--195 km s$^{-1}$. Although this is approximately half the expansion velocity calculated from the inner shell size relative to the time of eruption, it is still consistent with typical outflow velocities of red novae \citep{kaminski2018,steinmetz2024}. Therefore, we believe that the dust expansion is driven by stellar outflows typical of red novae.

Our 2010 model yields a total dust mass of 3.44\standform{-3} M\solar, almost an order of magnitude smaller than in our {\tt cd1ol} model. This indicates that dust condensation has continued several years after the merger event, and the increase in the inner shell radius indicates that the dust continues to be driven outwards from the central star. The average dust mass-loss rate is $\approx$7$\times$10$^{-4}$ M\solar\  yr$^{-1}$, which is unusually large for evolved stars but not unheard of in red nova remnants \citep[cf.][]{kaminski2021v838mon,TylendaKeck}. Our dust mass estimates are two orders of magnitude higher than the lower limit given by T13. The lack of expansion of the inner radius ($r_{\rm in}$) may indicate that dust is forming in close proximity to the central star, replenishing the inner shell with dust before it is driven outwards by the stellar outflows. This is consistent with a dust-driven wind in a low-gravity cool star, an analogue to winds of AGB stars. 

To illustrate the evolution of BLG-360 during the time between the two epochs considered so far, we extracted photometric measurements for the period 2011--2017. These include fluxes from WISE and Spitzer observations taken in 2015 and $JHK_s$ fluxes from the VISTA Variables in V\'ia Lactea survey \citep[VVV][]{minniti2010VVV} from 2015. The $H$ band flux was extracted from 23/08/2015 and the $K_s$ flux was taken from 07/08/2015. These fluxes are shown in Fig. \ref{fig: t13 + spitzer + obs}. $J$ band photometry was available but was not observed in 2015. We also extracted VVV photometry in the $K_s$ band between 2011--2017, which was resampled into bins separated by $\geq$ 30 days and converted to flux using the SVO filter service. The VVV source had coordinates of $\alpha$=17$^h$ 57$^m$ 38\fs98, $\delta$=--29$\degree$46\arcmin 05\farcs12, approximately consistent with the fitted coordinates from the ACA and SMA data. The subsequent light curve for VVV, as well as WISE, is shown in Fig. \ref{fig: neowise lc}, and shows a steady decline at 2.15, 3.40, and 4.60 \micron. The sources of the data are provided in Appendix \ref{appendix: intermediate phot}. The observations in Figs. \ref{fig: t13 + spitzer + obs} and \ref{fig: neowise lc} suggest smooth evolution of the object between the 2010 and 2017--2023 SED epochs. Since only NIR/IR data are available, the evolution of the object at longer wavelengths is essentially unknown.

The evolution of BLG-360 bears some resemblance to that of the extragalactic red nova M31-LRN-2015 \citep{blagorodnova2020}, which showed several epochs of dust formation up to $\approx$2.5 yrs after the eruption. The photometric evolution of BLG-360 at NIR wavelengths shown in Fig. \ref{fig: neowise lc} indicates rapid obscuration of the remnant immediately after the eruption, as well as late-time epochs of dust formation that leads to the dust becoming optically thick ($\tau_{\rm V}\sim\ $40). The ongoing dust production in BLG-360 also bears similarities to what has been observed in post-outburst evolution of the Galactic red nova V838 Mon, where the ongoing dusty wind \citep{kaminski2021v838mon,kaminskiLithium} is associated with circumstellar SiO masers \citep{ortiz2020}.
A sensitive search of SiO masers is thus highly desired in BLG-360. Even only the systemic velocity of the maser can help better constrain the distance and thus basic properties of this dusty remnant.

\section{The anomaly of BLG-360}\label{sect: compare transients}
BLG-360 stands out among red nova remnants because its sub-mm component was easily detected with the relatively small millimetre-wave arrays. Other red nova remnants at similar distances and progenitor masses did not produce enough dust for a direct sub-mm detection, including V1309 Sco and V4332 Sgr which were observed with the much more sensitive main ALMA array \citep{kaminski2018,steinmetz2024}. This suggests that the remnant has an unusually high dust-to-gas mass ratio. This ratio is, however, unknown, hampering any determinations of the total circumstellar mass. A question whether sources like BLG-360 are important dust producers in the context of the Galaxy dust budget and galaxy evolution is a question for another study.

Another remarkable feature of the BLG-360 remnant is the lack of spectroscopic signatures of gas, atomic or molecular, which are highly abundant in spectra of other known red-nova remnants of different ages \citep{kaminski2018,kaminski2021ckvul, steinmetz2024}. This lack of spectral lines does not only refer to the SMA and ACA data reported here, which covered the commonly-observed pure rotational lines of CO, but also to our $K_s$ band spectrum acquired at Gemini in 2015 (see Appendix \ref{appendix: Gemini spec}). The spectrum lacks any spectral features even though the first-overtone band of CO, H$_2$ lines, and multiple atomic transitions were covered. One way of explaining the paucity of gas features in spectra where continuum is detected at a high signal-to-noise is that the gas features are weak and difficult to detect at the expected high velocity dispersion of the ejecta and wind. Alternatively, the dust opacity could be so high that it blocks radiation from atomic and molecular transitions from gas located within the inner envelope. Indeed, the dust opacities of our best RADMC-3D model at 0.55 and 25 \micron\ are very high, 4918 and 318, respectively, but are quite modest at millimetre wavelengths (e.g. 1.6 at 1.3 mm). While the reason behind the missing gas lines is unclear, we can safely conclude that BLG-360 has become a very dusty remnant, making it a bright IR and (sub-)millimetre source.

The overall characteristics of BLG-360 presented in this paper make it stand out as a unique source, even among Galactic red novae. There are, however, several observational similarities between BLG-360 and other IR-bright eruptive stars. Multiple objects of differing classifications (other than red novae) have been discovered to have an IR component, including supernovae (SNe) \citep{myers2024} and SN imposters \citep{andrews2021}, young stellar objects (YSOs) \citep{habel2024}, and evolved stars \citep{nally2024}. The common explanation for the IR component in these sources is that, like BLG-360, they are hidden by optically thick dust shells or disks, and some should be considered infrared transients rather than optical ones.

There are several examples of IR-bright sources being considered as possible red nova candidates, such as N6946-BH1 \citep{kashi2017, beasor2024}, previously classified as a failed SN;  IRAS 19312+1950 \citep{qiu2023}, which exhibits properties of red novae and YSOs; and a millimetre ultra-broad-line object, G0.02467--0.0727, located near the Galactic Centre \citep{GinsburgMUBLO}. A relatively new class of IR transients, known as SPRITES \citep[eSPecially Red Intermediate-luminosity Transient Events;][]{kasliwal2017spirits}, includes one such object (SPIRITS 14azy) also identified as a possible red nova \citep{jencson2019spirits}.

To confirm the existence of red novae from their IR-bright remnants, we would require sub-mm and/or radio observations. Interferometric observations with ALMA, SMA, Very Large Array (VLA), and the upcoming Square Kilometer Array (SKA) may be able to observe atomic and molecular emission features common among red novae \citep{kaminski2018}, and resolve inner structures of the circumstellar environment that may detect evidence of bipolar outflows such as what we see for V838 Mon \citep{mobeen2024}, V4332 Sgr \citep{kaminski2018}, and V1309 Sco \citep{steinmetz2024}. The fact that red nova remnants, such as with BLG-360, can also be dominated by dust emission rather than by molecular emission and bipolar structures means that there may be other Galactic red novae present that have been overlooked due to their observable characteristics. A sub-mm or radio survey of IR-bright stellar sources may help to discover more red novae.
\section{Summary}\label{sect: summary}
We presented a full analysis and modelling of the optical-to-radio SED of the remnant OGLE-2002-BLG-360 about two decades after its red nova outburst. We used radiative transfer codes to constrain the properties of both the stellar remnant and its dusty circumstellar environment, including constraints on the geometry and dust composition. We found that:
\begin{enumerate}
    \item Based on models implemented using HYPERION and RADMC-3D, the circumstellar environment of BLG-360 can be approximated as spherically symmetric. The dust is concentrated within two shells, with a large cavity of no dust between. The dust is optically thick and shows no signatures of molecular and atomic gas usually seen surrounding other red nova remnants.
    \item We revise the distance of BLG-360 to 4.09 kpc, approximately half of the previous value of 8.2 kpc. The resulting luminosity of the source is 1646 L\solar. This is slightly dimmer than the modelled luminosity of the 2010 SED (2126 L\solar). This new distance adjusts the maximum luminosity of the eruption from 1.3\standform{4} to 3.2\standform{3} L\solar.
    \item Our best-fitting RADMC-3D models find that extended dust is necessary to reproduce the Rayleigh-Jeans tail, with the outer shell located at an order of $\approx$10$^4$ AU from the central star. Based on this distance and the time since the eruption, it is likely that the dust in the outer shell is likely formed from pre-merger material lost long before the 2002 eruption.
    \item Comparing our SED from 2017--2023 to data from 2010, we find that the stellar remnant has changed very little, with a constant radius and slightly cooler temperatures. Continued dust formation over the same period has increased the dust mass significantly, and this dust formed close to the star. The dust is driven outwards by active stellar winds at a velocity of 100--200 km s$^{-1}$.
    \item In our model, the dust is dominated by solid iron dust and olivine silicates. The large quantity of iron dust could be explained by an oxygen-rich AGB star, although this is not consistent with observations of the progenitor binary system.
\end{enumerate}
It is likely that other red nova remnants might also be fully obscured, and should be investigated further. Future observations with both the James Webb Space Telescope and sub-mm interferometers such as ALMA would be crucial in the identification of more red nova remnants that are enshrouded, similar to OGLE-2002-BLG-360. 
\begin{acknowledgements}
The authors would like to thank M. Lewis for their help in estimating the interstellar extinction. T.S would also like to thank S. Goldman for helpful discussions on SED modelling of deeply embedded objects. ThS and TK acknowledge funding from grant no 2018/30/E/ST9/00398 from the Polish National Science Center. N. B. acknowledges to be funded by the European Union (ERC, CET-3PO, 101042610). Views and opinions expressed are however those of the author(s) only and do not necessarily reflect those of the European Union or the European Research Council Executive Agency. Neither the European Union nor the granting authority can be held responsible for them. This paper makes use of the following ALMA data: ADS/JAO.ALMA\#2021.2.00017.S. ALMA is a partnership of ESO (representing its member states), NSF (USA) and NINS (Japan), together with NRC (Canada), MOST and ASIAA (Taiwan), and KASI (Republic of Korea), in cooperation with the Republic of Chile. The Joint ALMA Observatory is operated by ESO, AUI/NRAO and NAOJ. The Submillimeter Array is a joint project between the Smithsonian Astrophysical Observatory and the Academia Sinica Institute of Astronomy and Astrophysics and is funded by the Smithsonian Institution and the Academia Sinica. This work is based in part on observations made with the NASA/DLR Stratospheric Observatory for Infrared Astronomy (SOFIA). SOFIA is jointly operated by the Universities Space Research Association, Inc. (USRA), under NASA contract NNA17BF53C, and the Deutsches SOFIA Institut (DSI) under DLR contract 50 OK 2002 to the University of Stuttgart. All (SOFIA) observations were obtained with the German proposal 09-0088, PI. K. M. Menten. This publication makes use of data products from the Near-Earth Object Wide-field Infrared Survey Explorer (NEOWISE), which is a joint project of the Jet Propulsion Laboratory/California Institute of Technology and the University of Arizona. NEOWISE is funded by the National Aeronautics and Space Administration. This research is based in part on data collected at the Subaru Telescope, which is operated by the National Astronomical Observatory of Japan. 
Some of the data presented herein were obtained at Keck Observatory, which is a private 501(c)3 non-profit organization operated as a scientific partnership among the California Institute of Technology, the University of California, and the National Aeronautics and Space Administration. The Observatory was made possible by the generous financial support of the W. M. Keck Foundation. We are honoured and grateful for the opportunity of observing the Universe from Mauna Kea, which has the cultural, historical, and natural significance in Hawaii. The National Radio Astronomy Observatory is a facility of the National Science Foundation operated under cooperative agreement by Associated Universities, Inc. This scientific work uses data obtained from Inyarrimanha Ilgari Bundara / the Murchison Radio-astronomy Observatory. We acknowledge the Wajarri Yamaji People as the Traditional Owners and native title holders of the Observatory site. CSIRO’s ASKAP radio telescope is part of the Australia Telescope National Facility ({\url {https://ror.org/05qajvd42}}). Operation of ASKAP is funded by the Australian Government with support from the National Collaborative Research Infrastructure Strategy. ASKAP uses the resources of the Pawsey Supercomputing Research Centre. Establishment of ASKAP, Inyarrimanha Ilgari Bundara, the CSIRO Murchison Radio-astronomy Observatory and the Pawsey Supercomputing Research Centre are initiatives of the Australian Government, with support from the Government of Western Australia and the Science and Industry Endowment Fund. Based on data products from VVV	Survey (programme ID 179.B-2002) and VVVX Survey (programme ID 198.B-2004) observations made with the VISTA telescope	at the	ESO	Paranal	Observatory. Based on data obtained from the ESO Science Archive	Facility with DOI {\url {https://doi.eso.org/10.18727/archive/67}} (VVV) and	{\url {https://doi.eso.org/10.18727/archive/68}} (VVVX). Based on observations within projects GS-2014A-Q-53 and GN-2021B-FT-208 obtained at the international Gemini Observatory, a program of NSF NOIRLab, which is managed by the Association of Universities for Research in Astronomy (AURA) under a cooperative agreement with the U.S. National Science Foundation on behalf of the Gemini Observatory partnership: the U.S. National Science Foundation (United States), National Research Council (Canada), Agencia Nacional de Investigaci\'{o}n y Desarrollo (Chile), Ministerio de Ciencia, Tecnolog\'{i}a e Innovaci\'{o}n (Argentina), Minist\'{e}rio da Ci\^{e}ncia, Tecnologia, Inova\c{c}\~{o}es e Comunica\c{c}\~{o}es (Brazil), and Korea Astronomy and Space Science Institute (Republic of Korea). The national facility capability for SkyMapper has been funded through ARC LIEF grant LE130100104 from the Australian Research Council, awarded to the University of Sydney, the Australian National University, Swinburne University of Technology, the University of Queensland, the University of Western Australia, the University of Melbourne, Curtin University of Technology, Monash University and the Australian Astronomical Observatory. SkyMapper is owned and operated by The Australian National University's Research School of Astronomy and Astrophysics. The survey data were processed and provided by the SkyMapper Team at ANU. The SkyMapper node of the All-Sky Virtual Observatory (ASVO) is hosted at the National Computational Infrastructure (NCI). Development and support of the SkyMapper node of the ASVO has been funded in part by Astronomy Australia Limited (AAL) and the Australian Government through the Commonwealth's Education Investment Fund (EIF) and National Collaborative Research Infrastructure Strategy (NCRIS), particularly the National eResearch Collaboration Tools and Resources (NeCTAR) and the Australian National Data Service Projects (ANDS). Based on data obtained from the ESO Science Archive Facility with DOI: https://doi.org/10.18727/archive/68.
\end{acknowledgements}
\bibliographystyle{aa}
\bibliography{main.bib}
\begin{appendix}
\section{DUSTY}\label{appendix: dusty}
Here we describe our attempts to model the SED of BLG-360 with the 1-D radiative transfer code DUSTY \citep{dustyv2}. We based our initial model  on the analysis of the latest SED epoch (2010) considered by T13, which was also based on the assumption that the circumstellar dust was spherically symmetric. Then, we manually adjusted the parameters to better reproduce the observations. As the output SED of DUSTY is dimensionless, we scaled the model SED fluxes to the maximum value of our observed SED. The DUSTY input parameters include stellar temperature, dust density distribution, grain size distribution and composition (opacity), and dust temperature at the inner boundary. All other parameters are provided via scaling relations within the code.

\begin{figure}[h!]
    \centering\small
    \includegraphics[trim={40 0 0 0},width=\columnwidth]{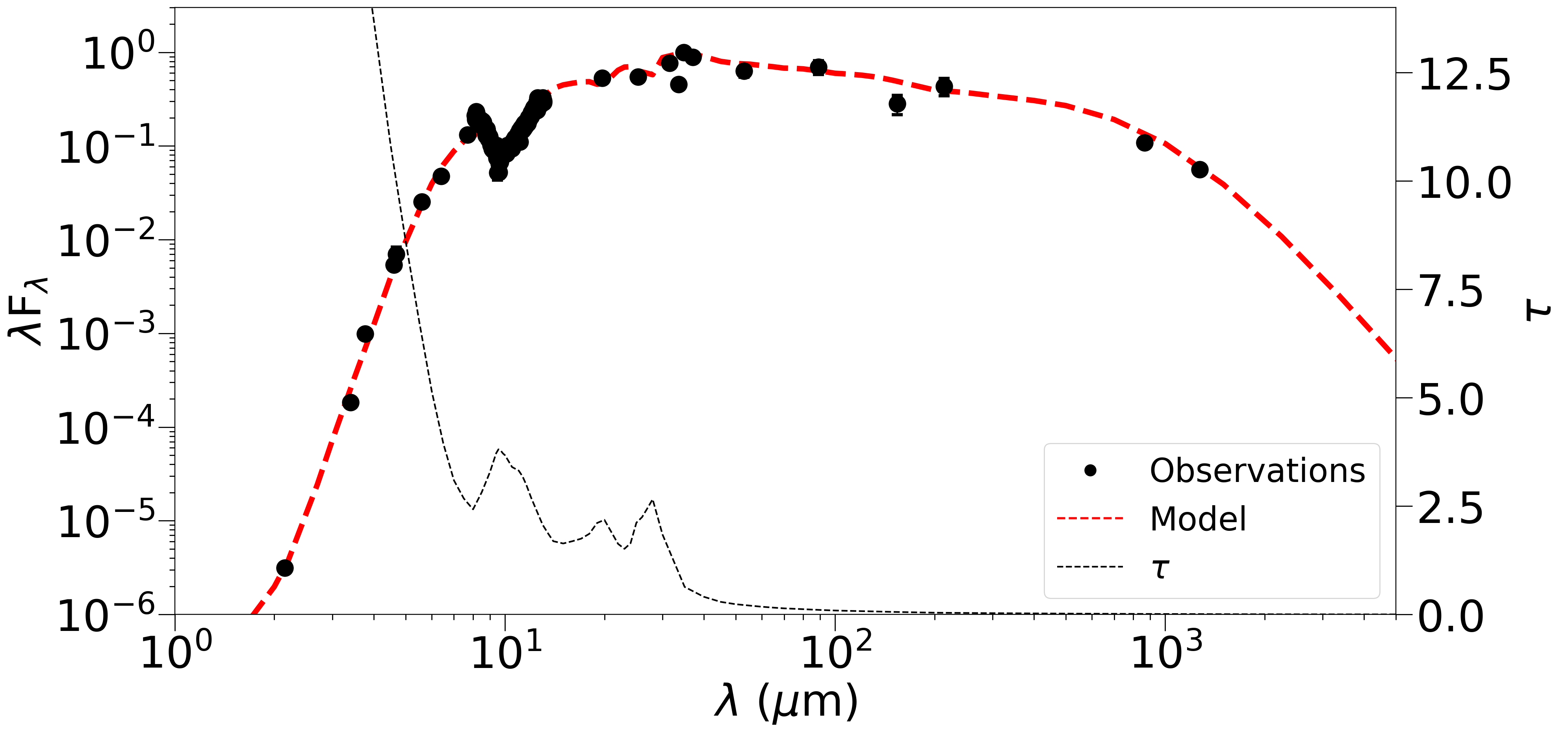}
    \caption{Best-fitting DUSTY model (red) overplotted with the observational SED, corrected for interstellar extinction. The black dashed line shows the opacity curve for this model.}
    \label{fig: model D}
\end{figure}

We present our best fitting model in Fig. \ref{fig: model D} and the model parameters in Table \ref{tab: model D}. The references for each grain type considered in our model are also shown in Table \ref{tab: model D}.  We see an oxygen-rich chemistry dominated by silicates, solid iron dust and MgS, although FeO is also needed to replicate the high NIR opacities;  alumina dust is necessary to reproduce the 10 \micron\ absorption profile, as without a contribution from alumina dust the absorption profile is too narrow. The star appears to be a K-type supergiant star with a radius of 661 R\solar, surrounded by a single spherical shell of dust where the density $\eta$ falls as $\eta \propto$ r$^{-1.1}$, and the grain size ranges between 0.3 and 1.0 \micron. The inner boundary dust temperature is 300 K.

The key outcome from this model is that it requires the dust to extend out to 20 pc, which is on the scale of a giant molecular cloud and therefore unrealistic. This large dust content in the model is imposed by the relatively high submm-wave fluxes of BLG-360. Assuming a distance of 8.2 kpc (T13), a tangential distance of 20 pc is equivalent to 8\farcm4. In our ACA and SMA maps, we would see such an extended structure, unless the dust emission is sufficiently weak to not be detected. Note also, that the apertures used to retrieve the source fluxes were much smaller than that. W present the DUSTY result given the high level of accuracy in the fit and because it emphasizes the extremely dusty nature of the circumstellar environment of BLG-360.

\begin{table}[h!]
    \centering
    \caption{Table of physical parameters of the stellar remnant and dust shell found from DUSTY modelling.}
    \begin{tabular}{ccccccccc}\hline
    T$_{\rm eff}$ & R$_{\star}$ & r$_{in}$ & r$_{out}$ & T$_d^{in}$ & a$_{min}$ & a$_{max}$ & q & p\\
    (K) & (R\solar) & (AU) & (pc) & (K) & (\micron) & (\micron) & & \\\hline
    4500 & 661 & 208 & 20 & 300 & 0.30 & 1.00 & 2 & --1.1\\
    \end{tabular}
    \begin{tabular}{ccc}\hline
    Grain type & Composition& Reference\\\hline
    astro-silicates & 30\% & \citet{drainelee1984}\\
    FeO & 10\% & \citet{henning1995}\\
    Al$_2$O$_3$ & 10\% & \citet{begemann1997} \\
    \multirow{2}{*}{Metallic Fe} & \multirow{2}{*}{25\%} & \citet{pollack1994} \\
    & & \citet{henningstognienko1996} \\
    Mg$_{0.9}$Fe$_{0.1}$S & 25\% &  \citet{begemann1994} \\
    \hline
    \end{tabular}
    \tablefoot{$R_{\star}$ is the stellar radius, $r_{in,out}$ is the inner and outer radius of the dust shell, a is the grain size, q is the power law index of the grain size distribution, and p is the power law index of the density distribution.}
    \label{tab: model D}
\end{table}
\section{Gemini spectroscopy}\label{appendix: Gemini spec}
\begin{figure}
    \centering
    \includegraphics[trim=15 0 18 10, width=\columnwidth, clip]{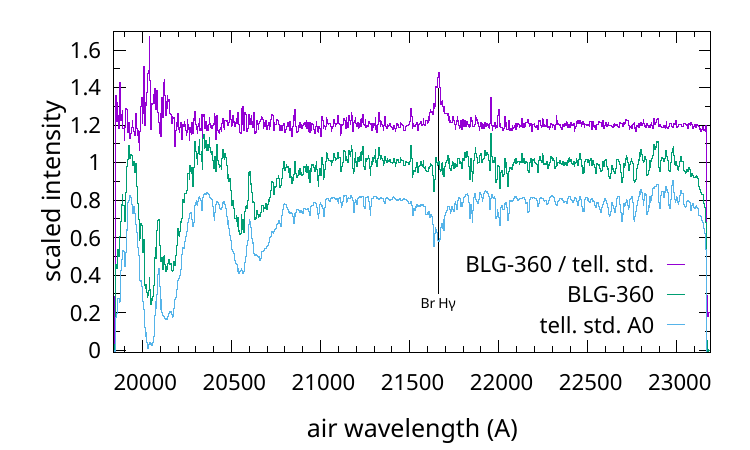}
    \caption{$K$ band Gemini spectrum of BLG-360 (green) is compared to a telluric standard (blue). The top plot shows the BLG-360 spectrum divided by that of the standard. The Bracket $\gamma$ line shows up because it is a feature in the standard's spectrum. No lines intrinsic to BLG-360 can be seen.} 
    \label{fig-gemini-spec}
\end{figure}

Long-slit spectroscopy was obtained with FLAMINGOS-2 on Gemini South on 20/02/2014 via program GS2014A-Q-53, in the “medium object” read mode. Four 150 s spectra were obtained using an ABBA jitter pattern. The $K_s$ grism and filter were used with the 2-pixel slit, producing a spectrum from 1.98 to 2.32 \micron. The A0V star HIP 88247, observed before the target and at similar airmass of $\approx$1.5, was used as a calibrator to remove telluric features. Flat fields and spectral arcs (Argon lamp) were obtained using lamps in the calibration unit. The data were reduced in the standard IRAF Gemini package, with flux calibration achieved by division by the Vega spectrum. As shown in Fig. \ref{fig-gemini-spec}, the spectrum of BLG-360 does not show any spectral features in the covered range.

\section{Intermediate epoch photometry}\label{appendix: intermediate phot}
Here we describe the intermediate photometry shown in Fig. \ref{fig: t13 + spitzer + obs}. BLG-360 was observed on 21/09/2015 in channels W1 and W2. Fluxes were extracted using the same method as described in Sect. \ref{sect: wise}. The obtained fluxes were 0.17$\pm$0.010 and 0.57$\pm$0.011 Jy for W1 and W2, respectively.

BLG-360 was also observed with Spitzer using the 3.6 and 4.5 \micron\ sub-arrays on the IRAC instrument. Two separate runs at each wavelength were observed on 17/07/2015 and again on 25/11/2015. Each run contains four stacked images, containing 64 individual exposures with integration times of 0.4 and 0.1 s for the 3.6 and 4.5 \micron\ sub-arrays, respectively. The data were processed with the Spitzer Science Center Pipeline vS19.2.0 and the stacked images were combined and averaged for each run. We performed aperture photometry using the \texttt{photutils} python package, and then averaged the measured fluxes together to obtain a single flux measurement for each wavelength array. We performed aperture corrections corresponding to a source radius of 5 pixels and a background annulus of 5--10 pixels. Our aperture radii varied between 5--7 pixels for the source, and was fixed at 4 pixels for the background. As BLG-360 lies within a crowded field, we estimated the background using a separate aperture centered on a region of blank sky in the images rather than an annulus centered on the source. We also performed colour corrections assuming a power-law spectrum (F$_{\nu}\ \propto\ \nu^{-\alpha}$), where $\alpha$=--2.

BLG-360 was regularly observed in broadband $ZYJHK_s$ filters by the VISTA Variables in V\'ia Lactea survey \citep[VVV][]{minniti2010VVV} survey. We extracted $JHK_s$ magnitudes using the ESO catalogue facility \footnote{https://www.eso.org/qi/} from the VIRAC2 time series catalogue \citep{Smith}. No $ZY$ magnitudes were available for our source. We searched for detected sources within a 0\farcs5 search cone centred on the coordinates of BLG-360, which provided a single source. We then extracted the $JHK_s$ magnitudes for specific dates: the $J$ band was observed on 31/08/2010 (the latest available date for BLG-360 in this filter), the $H$ band was observed on 23/8/2015, and the $K_s$ band was observed on 07/08/2015. We converted them to fluxes using the Vega zero-point system. Central wavelengths and zero-point fluxes were taken from SVO. 

All intermediate fluxes are shown in Table \ref{tab: intermediate fluxes}.
\begin{table}
    \caption{Fluxes from WISE, Spitzer, and VVV for BLG-360 for the intermediate epoch between the observation epoch for the T13 SED and our own 2017--2023 SED.}
    \centering\normalsize
    \begin{tabular}{cccc}\hline
Facility & Filter & $\lambda_c$ (\micron) & Flux (Jy)\\\hline
VVV & $J$ & 1.25 & (2.06$\pm$1.35)\standform{-4} \\
VVV & $H$ & 1.65 & (1.72$\pm$0.60)\standform{-4} \\
VVV & $K_s$ & 2.15 & (2.77$\pm$0.11)\standform{-3} \\
WISE & W1 & 3.4 & 0.17$\pm$0.010 \\
WISE & W2 & 4.6 & 0.57$\pm$0.011 \\
Spitzer & 3.6 & 3.6 & 0.13 \\
Spitzer & 4.5 & 4.5 & 0.14 \\\hline
    \end{tabular}
    \label{tab: intermediate fluxes}
\end{table}
\section{Alternative photometry extraction}\label{appendix: burris photometry}
\begin{figure}[h!]
    \centering
    \includegraphics[trim=0 0 0 0, width=\columnwidth]{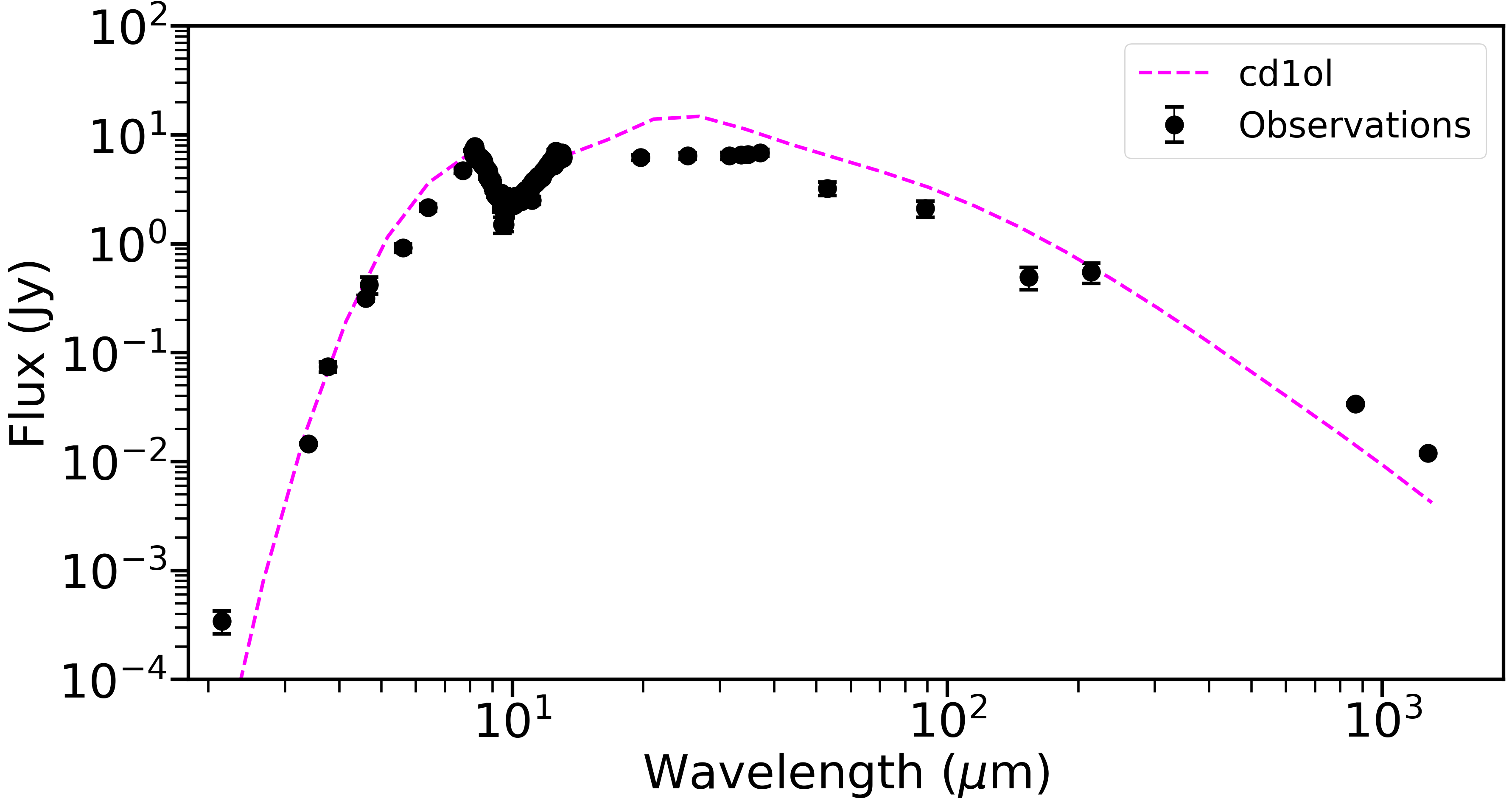}
    \caption{Our SED containing the PSF-corrected FORCAST photometry, along with the best-fitting \texttt{cd1ol} model in magenta.} 
    \label{fig: psf forcast}
\end{figure}

To examine whether the fluctuations in the FORCAST fluxes are real features, we extracted fluxes following the PSF-matching photometry routine described in \citet{burris2023}. The flux extraction for the F077 filter failed as the PSFs were too unstable. We present the alternative fluxes in Table \ref{tab: psf forcast} and show the SED plotted along with the \texttt{cd1ol} model in Fig. \ref{fig: psf forcast}. The new fluxes show a steady rise between 5--19 \micron, and then a flat SED between 19--37 \micron. The PSF photometry shows that the fluctuations are due to noise within the large FORCAST photometry extraction aperture. Despite the different photometry, we find that the \texttt{cd1ol} model is still the best-fitting model.
\begin{table}[h!]
    \caption{FORCAST fluxes extracted using the standard photometry routine and after PSF variation corrections.}
    \centering\normalsize
    \begin{tabular}{ccc}\hline
Wavelength & Original flux (Jy) & PSF-extracted flux (Jy)\\\hline
5.61 & 1.22$\pm$0.09 & 0.89$\pm$0.08\\
6.35 & 2.00$\pm$0.13 & 2.10$\pm$0.16\\
8.64 & 4.68$\pm$0.28 & 4.30$\pm$0.26\\
11.01 & 2.69$\pm$0.18 & 2.48$\pm$0.21\\
19.70 & 7.29$\pm$0.44 & 6.15$\pm$0.38\\
25.24 & 5.81$\pm$0.37 & 6.41$\pm$0.44\\
31.36 & 6.58$\pm$0.44 & 6.40$\pm$0.51\\
33.57 & 3.65$\pm$0.41 & 6.50$\pm$0.55\\
34.64 & 7.76$\pm$0.52 & 6.57$\pm$0.52\\
36.98 & 6.44$\pm$0.43 & 6.85$\pm$0.51\\\hline
    \end{tabular}
    \label{tab: psf forcast}
\end{table}
\end{appendix}
\end{document}